# Investigating the efficiency of the Asian handicap football betting market with ratings and Bayesian networks

Anthony C. Constantinou[1,2]

1. Bayesian Artificial Intelligence research lab, Risk and Information Management (RIM) research group, School of Electronic Engineering and Computer Science, Queen Mary University of London (QMUL), London, UK, E1 4NS.
   E-mail: a.constantinou@qmul.ac.uk
2. The Alan Turing Institute, UK.

**ABSTRACT:** Despite the massive popularity of the Asian Handicap (AH) football (soccer) betting market, its efficiency has not been adequately studied by the relevant literature. This paper combines rating systems with Bayesian networks and presents the first published model specifically developed for prediction and assessment of the efficiency of the AH betting market. The results are based on 13 English Premier League seasons and are compared to the traditional market, where the bets are for win, lose or draw. Different betting situations have been examined including a) both average and maximum (best available) market odds, b) all possible betting decision thresholds between predicted and published odds, c) optimisations for both return-on-investment and profit, and d) simple stake adjustments to investigate how the variance of returns changes when targeting equivalent profit in both traditional and AH markets. While the AH market is found to share the inefficiencies of the traditional market, the findings reveal both interesting differences as well as similarities between the two.

*Keywords*: directed acyclic graph, football prediction, graphical models, profitability, rating system, return-on-investment, soccer prediction.

## 1. Introduction

Football prediction models have become immensely popular over the last couple of decades, and this is due to the increasing popularity of football betting. In the academic literature, such models typically focus on predicting the outcome of a match in terms of home win, draw, or away win; known as the 1X2 distribution. Several types of models have been published for this purpose and include rating systems (Leitner et al., 2008; Hvattum & Arntzen, 2010; Constantinou & Fenton, 2013; Wunderlich & Memmert, 2018), statistical methods (Dixon & Coles, 1997; Rue & Salvesen, 2000; Crowder et al., 2002; Goddard, 2005; Angelini & De Angelis, 2017), machine learning techniques (Huang & Chang, 2010; Arabzad et al., 2014; Pena, 2014), knowledge-based systems (Joseph et al., 2006), and hybrid methods that combine any of the above (Constantinou & Fenton, 2017; Constantinou, 2018; Hubacek et al., 2018). In the recent special issue international competition *Machine Learning for Soccer*, the models that topped the performance table are hybrid and heavily rely on rating systems (Constantinou, 2018; Hubacek et al., 2018).

In Asia, the most popular form of betting (also common in Europe) is the so-called Asian Handicap (AH). This form of betting introduces a hypothetical handicap (i.e., advantage) typically in favour of the weaker team. Specifically, traditional AH introduces a goal deficit to the team more likely to win before kick-off. The manipulation of the match outcome creates interesting situations in which betting is determined by hypothetical, rather than actual, match outcome. Examples of the various types of AH betting are provided in Section 2. This type of betting has also become popular in the UK over the last couple of decades. Football (soccer) syndicates are rumoured to bet millions per week, often on behalf of clients, on AH outcomes offered by bookmakers in Asia (Williams-Grut, 2016). This is because the Asian markets attract higher volumes of bets and offer greater market liquidity. Estimates suggest that over 70% of the betting turnover for football is recorded with Asian bookmakers (Kerr, 2018).

Whereas there is a large literature analysing traditional betting strategies, with researchers investigating how to optimise their betting, there has been only four published papers involving some analysis related to AH. Specifically, Vlastakis et al (2008) used AH odds as one of their model variables to predict match scores and showed that they are a strong predictor of match outcomes. Grant et al (2018) used AH odds, in conjunction with 1X2 odds, to analyse arbitrage opportunities and showed that these exist across a large number of fixed-odds and exchange market odds. Hofer and Leitner (2017) described how to derive information from live AH and Under/Over odds in order to maximise expected returns. Finally, and in an effort to educate gamblers, Hassanniakalager and Newall (n.d.) investigated the product risk associated with different football odds and showed that the AH odds would generally generate lower losses compared to other popular types of bet such as the 1X2, Under/Over, and correct score. Remarkably, no previous published work involves a model specifically designed for, and assessed with, AH bets.

The purpose of this paper is to investigate the efficiency of the AH market in relation to the 1X2 market. The 1X2 market has been extensively studied and the literature provides mixed empirical evidence regarding its efficiency, with most evidence pointing towards a weak-form efficient market (Giovanni & De Angelis, 2019). In this paper, the efficiency of both markets is measured in terms of the ability of the model in discovering profitable betting opportunities given both average and maximum market odds. The model is based on a modified version of the pi-rating system, which is a previously published football rating system (Constantinou & Fenton, 2013), that generates ratings that reflect team scoring ability. The ratings are provided as input to a novel hybrid Bayesian Network (BN) model specifically constructed to simulate the influential relationships between possession, shots, and goals, to predict both 1X2 and AH outcomes.

A BN is a type of a probabilistic model introduced by Pearl (Pearl, 1985) that consists of nodes and arcs. Nodes represent variables and arcs represent conditional dependencies. A BN that consists of both discrete and continuous variables, such as the one constructed in this study, is called a hybrid BN. Each variable has a corresponding Conditional Probability Distribution (CPD) that captures the magnitude as well as the shape of the relationship between directly linked variables. If we assume that the arcs in the BN represent influential relationships, then such a model can be viewed as a causal graph and represents a unique Directed Acyclic Graph (DAG) that can be used for interventional analysis. Otherwise, the arcs represent conditional dependencies that are not necessarily causal relationships, and such a BN is not a unique DAG but rather a Partial DAG that represents an equivalence class of DAGs. For a quick introduction to BNs, with a focus on football examples, see (Constantinou & Fenton, 2018).

Based on 13 English Premier League seasons and betting simulations under different assumptions, the findings reveal interesting differences as well as similarities between the AH and 1X2 markets. Importantly, the AH market is found to share inefficiencies with the traditional 1X2 market, and this provides opportunities for 'beating' the market. The paper is structured as follows: Section 2 describes the rules of the AH betting market, Section 3 describes the model, Section 4 covers the data and the process of model fitting, Section 5 presents the results, and Section 6 provides the concluding remarks.

## 2. Asian Handicap betting rules

In what follows, the *decimal* odds system is used (also known as European odds) for payoff in the event of winning a bet. The decimal odds represent the total return ratio of the stake; implying that the stake is already included in the decimal number. For example, a payoff of '3' returns three times the stake; i.e., a bet of £1 would return 1×3=£3 (£2 profit). Odds also reflect probability that incorporates the bookmakers' profit margin. An example from data is the Arsenal versus Crystal Palace match played on 21/04/2019 with average 1X2 market odds $\{1.54, 4.44, 6.03\}$, corresponding to probabilities $\{64.94\%, 22.52\%, 16.58\%\}$. Summing up these probabilities gives us $104.04\%$, and the implied average profit margin of $4.04\%$.

The AH betting market operates such that adversaries are handicapped according to their difference in strength. The term handicap means that one team is assigned a hypothetical score (including fractional) advantage before the match is played. All types of AH betting offered by the bookmakers involve two possible outcomes. This means that the AH betting market reduces the possible match outcomes from three (i.e., 1X2) to two. Each binary outcome corresponds to each team winning, with the odds adjusted subject to the given handicap.

The standard AH betting involves assigning a hypothetical score advantage to the underdog. This represents the most common type of AH betting, and aims to make the contest equal. That is, the handicap applied is the one[*] that optimises the odds, for both teams to win, towards 2 (or 50% chance of winning) and hence, it maximises the uniformity of the AH payoff distribution. While this represents the most popular type of bet, the AH market offers different types of score advantage, each of which we discuss below, including the possibility to assign a hypothetical score advantage to the favourite rather than the underdog.

The market offers three types of AH betting that need to be modelled explicitly into the model, as well as in the betting simulation. These are:

i. *Whole goal handicap*: A team is given a whole-goal handicap such as -1 or +1. In this case, the possibility of a draw is eliminated by removing the draw outcome from the equation and normalising the probabilities of the residual two outcomes to sum up to 1. If a handicap draw is observed, the bet is voided (refunded).

   Let us revisit the example from data discussed at the beginning of this section, involving Arsenal versus Crystal Palace with average 1X2 market odds $\{1.54, 4.44, 6.03\}$. Arsenal was the strong favourite. The bookmakers introduced the handicap of -1, which maximised the uniformity of the AH distribution with odds $\{1.87, 1.99\}$. The match ended 2-3; i.e., -1 for Arsenal. The AH winner was Crystal Palace since it won the match by one goal difference, which makes it two goals difference given the handicap; i.e., this made the settlement score, which is the match result after the handicap is considered, equal to 2. Table 1 illustrates how the whole-goal AH is determined based on other hypothetical score lines between Arsenal and Crystal Palace.

[*] The other handicaps do not share the same market liquidity; implying limited stakes and possibly also higher profit margins, for the bookmaker, due to lower competition.

**Table 1.** The whole-goal AH outcome for different hypothetical scores between Arsenal and Crystal Palace.

| Arsenal goals | Crystal Palace goals | Score difference | Handicap | Settlement score | AH winner |
|---|---|---|---|---|---|
| 1 | 0 | 1 | -1 | 0 | Void |
| 1 | 1 | 0 | -1 | -1 | Crystal Palace |
| 3 | 1 | 2 | -1 | 1 | Arsenal |
| 4 | 1 | 3 | -1 | 2 | Arsenal |
| 0 | 0 | 0 | -1 | -1 | Crystal Palace |
| 0 | 1 | -1 | -1 | -2 | Crystal Palace |
| 1 | 2 | -1 | -1 | -2 | Crystal Palace |

ii. *Half-goal handicap*: A team is given a half-goal handicap such as -1.5 or +1.5. Assuming a match between $X$ and $Y$ and a handicap of +1.5 (i.e., $X$ receives a 1.5 goal advantage), and that a bet is placed on $X$, the bet would win as long $X$ does not lose by more than one goal difference; otherwise the bet is lost. In this case, the possibility of a draw is eliminated by the handicap itself, since it is not possible for the settlement score to be a draw.

An example from data is the Liverpool versus Wolves match played on 12/05/2019 with average 1X2 market odds $\{1.30, 5.62, 10.17\}$. Liverpool was the strong favourite. The bookmakers introduced the handicap of -1.5, which maximised the uniformity of the AH distribution with odds $\{1.91, 1.95\}$. The match ended 2-0 (i.e., +2) in favour of Liverpool. The AH winner was Liverpool since it won the match by two goals difference; i.e., 0.5 goals more than the handicap. This made the settlement score equal to 0.5. Table 2 illustrates how the half-goal AH is determined based on other hypothetical score lines between Liverpool and Wolves.

**Table 2.** The half-goal AH outcome for different hypothetical scores between Liverpool and Wolves.

| Liverpool goals | Wolves goals | Score difference | Handicap | Settlement score | AH winner |
|---|---|---|---|---|---|
| 1 | 0 | 1 | -1.5 | -0.5 | Wolves |
| 1 | 1 | 0 | -1.5 | -1.5 | Wolves |
| 3 | 1 | 2 | -1.5 | 0.5 | Liverpool |
| 4 | 1 | 3 | -1.5 | 1.5 | Liverpool |
| 0 | 0 | 0 | -1.5 | -1.5 | Wolves |
| 0 | 1 | -1 | -1.5 | -2.5 | Wolves |
| 1 | 2 | -1 | -1.5 | -2.5 | Wolves |

iii. *Quarter-goal handicap*: A team is given a quarter-goal handicap such as -0.25 or +0.25. This type of handicap is, in fact, a combined whole-goal and a half-goal handicap. For example, if we bet £10 on the away team to win given AH -0.25 with odds 2 (i.e., 50%), the stake would be divided between the nearest whole-goal and half-goal handicaps. That is, a £5 bet will be placed on the away team to win given AH ±0[†] with odds ~2.5 (i.e., 40%) and another £5 bet on the away team to win given AH -0.5 with odds ~1.66 (i.e., 60%). Note that the odds for the quarter-goal handicap reflect the average payoff, in terms of probability, of the two nearest handicaps. Since this is a combination of two bets, each bet is executed independently. For example, a score of 0-0 would have resulted in voiding AH ±0 (i.e., £5 are returned) and winning AH -0.5 (i.e., £5×1.66=£8.3 are returned).

[†] A zero-goal AH implies no handicap, but that there must be a match winner; otherwise, the bet is voided.

An example from data is the Fulham versus Newcastle match played on 12/05/2019 with average 1X2 market odds $\{2.50, 3.53, 2.78\}$. Fulham was the weak favourite. The bookmakers introduced the handicap of -0.25, which maximised the uniformity of the AH distribution with odds $\{2.15, 1.75\}$. The match ended 0-4(i.e., -4) in favour of Newcastle. The AH winner was Newcastle, since it won the match by four goals difference; i.e., 4.25 goals more than the handicap. This made the settlement score equal to -4.25. Table 3 illustrates how the quarter-goal AH is determined based on other hypothetical score lines between Fulham and Newcastle.

**Table 3.** The quarter-goal AH outcome for different hypothetical scores between Fulham and Newcastle.

| Fulham Goals | Newcastle goals | Score difference | Handicap | Settlement score | AH winner |
|---|---|---|---|---|---|
| 1 | 0 | 1 | -0.25 (0 and -0.5) | 1 and 0.5 | Fulham |
| 1 | 1 | 0 | -0.25 (0 and -0.5) | 0 and -0.5 | Void and Newcastle |
| 3 | 1 | 2 | -0.25 (0 and -0.5) | 2 and 1.5 | Fulham |
| 4 | 1 | 3 | -0.25 (0 and -0.5) | 3 and 2.5 | Fulham |
| 0 | 0 | 0 | -0.25 (0 and -0.5) | 0 and -0.5 | Void and Newcastle |
| 0 | 1 | -1 | -0.25 (0 and -0.5) | -1 and -1.5 | Newcastle |
| 1 | 2 | -1 | -0.25 (0 and -0.5) | -1 and -1.5 | Newcastle |

## 3. The model

The overall model combines ratings with BNs. The rating system captures the skill of teams over time, and provides the ratings as an input into the BN model which captures the magnitude of the relationships between variables of interest. The two subsections that follow describe the rating system and the BN model respectively.

### *3.1. The rating system*

The pi-rating is a football rating system that determines team ability based on the relative discrepancies in scores between adversaries. It was first introduced in (Constantinou & Fenton, 2013) and thereafter used in (Constantinou, 2018; Hubacek et al., 2018; 2019; Van Cutsem, 2019; Wheatcroft, 2020). Modified versions of the pi-rating also formed part of the top two performing models in the international competition *Machine Learning for Soccer* (Constantinou, 2018; Hubacek et al., 2018). This paper makes use of the original pi-rating system (Constantinou & Fenton, 2013), with two modifications described below.

The pi-ratings assign a 'home' ($H$) and an 'away' ($A$) rating to each team, to account for team-specific home advantage and away disadvantage. Therefore, when a team $X$ plays against team $Y$, the match prediction is determined by team's $X$ rating $H$ versus team's $Y$ rating $A$. The ratings are revised after each match based on two learning rates: a) the learning rate $\lambda$ which determines to what extent new match results override previous match results in terms of the impact in determining current team ratings, and b) the learning rate $\gamma$ which determines to what extent performances at home grounds influence a team's away rating and vice versa. Therefore, at the end of a match between teams $X$ and $Y$, the new ratings at time $t$ are revised given the most recent ratings at time $t-1$ as follows:

$X$'s $H$ rating: $R_{XH}^{t} = R_{XH}^{t-1} + e_H\lambda$

$X$'s $A$ rating: $R_{XA}^{t} = R_{XA}^{t-1} + \gamma(R_{XH}^{t} - R_{XH}^{t-1})$

$Y$'s $A$ rating: $R_{YA}^{t} = R_{YA}^{t-1} + e_A \lambda$

$Y$'s $H$ rating: $R_{YH}^{t} = R_{YH}^{t-1} + \gamma(R_{YA}^{t} - R_{YA}^{t-1})$

where $e$ is the error between the observed goal difference $\Delta_o$ and rating difference $\Delta_p$ which, for home and away teams, is measured as follows:

$$e_H = \Delta_{oH} - \Delta_{pH} \qquad \text{and} \qquad e_A = \Delta_{oA} - \Delta_{pA}$$

respectively, where

$$\Delta_{oH} = G_{oH} - G_{oA} \qquad \text{and} \qquad \Delta_{oA} = G_{oA} - G_{oH}$$

$$\Delta_{pH} = G_{pH} - G_{pA} \qquad \text{and} \qquad \Delta_{pA} = G_{pA} - G_{pH}$$

where $G_{oH}$ and $G_{oA}$ are goals observed for home and away teams respectively, and similarly $G_{pH}$ and $G_{pA}$ are goals predicted for home and away teams.

While the original pi-ratings represent a diminished expectation of goal difference against the average opponent in the data, in this paper they represent the actual goal difference expectation. Specifically, the rating equation in this paper is simplified not to include the deterministic function $\Psi(\mathrm{e}) = 3 \times \log_{10}(1 + \mathrm{e})$ defined in the original paper (Constantinou & Fenton, 2013), which is a function that diminishes the importance of each additional goal difference under the assumption that a win is more important than increasing goal difference. The justification for this first modification is that, in AH, we are only interested in goal differences and thus, the motivation here is to optimise for goal difference rather than the ability to win matches.

The second modification involves the learning rate λ. In this paper, λ is multiplied by $k$ when a match involves at least one team which had previously played less than 38 matches, according to available data. This modification aims to increase the speed by which team ratings converge for new teams during their first EPL season (each team plays 38 matches in a season), and is expected to be especially impactful during the very first season in the data since, at that point, all teams are considered 'new' by the rating. Therefore, the revised pi-ratings exclude $\Psi(e)$, defined above, and include $k$, as follows:

$$R_{XH}^{t} = R_{XH}^{t-1} + e_H \lambda k \qquad \text{and} \qquad R_{YA}^{t} = R_{YA}^{t-1} + e_A \lambda k$$

where $k = 3$ for match instances in which both teams had previously played less than 38 matches; otherwise $k = 1$. The parameter $k$ was optimised in terms of minimising prediction error $e$. A limitation here is that the $k$ parameter was optimised given integer inputs from 1 to 10. For future work, it is recommended that the $k$ parameter is optimised given real numbers.

### *3.2. The BN model*

The graph of a BN model can be automatically discovered from data, determined by knowledge, or a combination of the two. Learning the correct graph of a BN from data remains a major challenge in the fields of probabilistic machine learning and causal discovery. While some structure learning algorithms perform well with synthetic data, it is widely acknowledged that this level of performance does not extend to real-world data which typically incorporate noise and latent confounders.

In disciplines like bioinformatics, applying structure learning algorithms can reveal new insights that would otherwise remain unknown. However, these algorithms are less effective in areas with access to domain knowledge, such as in sports. As a result, the BN model in this paper has had its graphical structure determined by the temporal fact that possession influences the number of shots created, which in turn influence the number of shots on target, and which in turn influence the number of goals scored. Each of these factors is also dependent on the level of rating difference between the two teams, as illustrated in Fig 1. This type of model can also be characterised as a hierarchical Bayesian model of which the network is the structural representation, and where the nodes represent variables or hyperparameters of statistical distributions.

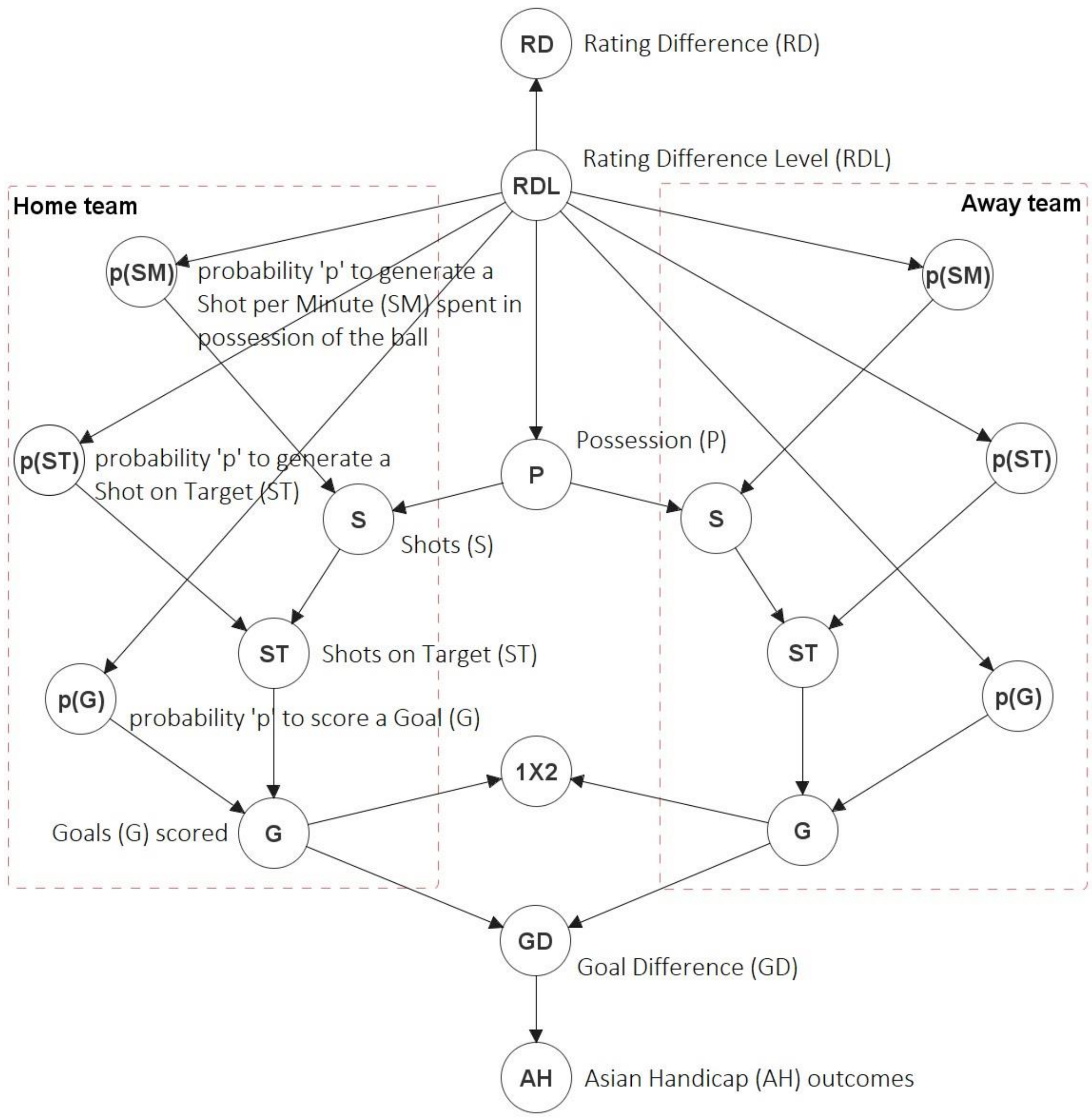

**Figure 1.** The BN model. The *Rating Difference (RD)* is the only observable node in the network, determined by the pi-ratings, and represents the difference between the home team's prior home rating and the away team's prior away rating $R_{XH}^{t-1} - R_{YA}^{t-1}$.

The temporal order of events in the BN graph naturally captures the importance of each event in predicting goals scored. For example, the graph assumes that shots on target have a direct influence on goals scored, whereas possession has an indirect influence and hence, while influential, it is assumed to be less impactful than shots on target. While the temporal order defines direct and indirect influences, note that the magnitude of direct influences is still determined by data.

For each match, the prior ratings are retrieved and the difference in team ratings is used as an input into the BN model, which is a Hybrid BN model consisting of both discrete and continuous variables, designed in AgenaRisk (Agena, 2019). Specifically, the actual input is the difference between prior home and away ratings, and is passed to the BN model as an observation to node *Rating Difference* ($RD$) in the form of

$$R_{XH}^{t-1} - R_{YA}^{t-1}$$

To ensure that the BN model is trained accurately with respect to the rating data, the parameterisation of the CPDs is also restricted to match instances in which both teams had previously played at least 38 matches. All the residual variables in the BN model are latent. Specifically,

i. The node $RD$, which represents the observable rating difference between teams, is a mixture of Gaussian probability density functions $\sim N(\mu, \sigma^2)$; one for each state of node *Rating Difference Level* ($RDL$). Specifically, for $-\infty < \text{RD} < \infty$,

$$f(RD|\mu, \sigma^2, RDL) = \left[ \left( \frac{1}{\sqrt{2\pi\sigma^2}} e^{-\frac{(RD-\mu)^2}{2\sigma^2}} \right) \middle| RDL \right]$$

where parent $RDL$ is a discrete distribution, $\mu$ is the average rating difference and $\sigma^2$ the variance of the rating differences. $RDL$ consists of 23 states[‡], where each state corresponds to a pre-determined level of rating difference as shown in Table 4. For example, the rating difference level 3 is parameterised based on all historical match instances in which adversaries had rating difference $RD = R_{XH}^{t-1} - R_{YA}^{t}$ ranging from 1.765 to <1.93. The granularity of the 23 states has been chosen to ensure that for any combination of rating difference there is enough data points (more than 50) for a reasonably well informed prior.

**Table 4.** Predetermined levels of rating difference.

| $RDL$ | 1 | 2 | 3 | … | 21 | 22 | 23 |
|---|---|---|---|---|---|---|---|
| $RD$ | ≥ 2.095 | ≥ 1.93 & < 2.095 | ≥ 1.765 & < 1.93 | intervals of 0.165 rating | ≥ −1.205 & < −1.04 | ≥ −1.37 & < −1.205 | < −1.37 |
| Data points | 55 | 68 | 107 | … | 93 | 58 | 55 |

[‡] The decision to discretise $RDL$ represents a practical choice for Hybrid BN modelling. In this case, discretising $RD$ into $RDL$ was necessary to capture conditional Beta-Binomial relationships from *Possession* to *Goals scored*, given the rating difference between adversaries.

ii. The node $P$, which represents ball possession, is a mixture of probability density functions $\sim Beta(a, \beta)$; one for each state of $RDL$. Specifically, for P ∈ $[0,1]$,

$$f(P|a, \beta, RDL) = \left( \frac{P^{a-1}(1-P)^{\beta-1}}{Beta(a, \beta)} \middle| RDL \right)$$

where $Beta(a, \beta)$ is the $Beta$ function, $a$ is the first shape parameter of the Beta function, also known as the *alpha* parameter, and represents the number of minutes the home team is in possession of the ball, and $\beta$ is the second shape parameter of the Beta function, also known as the *beta* parameter, that represents the number of minutes the away team is in possession of the ball. Thus, $P$ reflects the possession rate associated with the home team, over a Beta distribution, whereas for the possession of the away team the model assumes $1 - P$.

iii. The node $p(SM)$, which represents the probability to generate a shot per minute spent in possession of the ball, is also a mixture of probability density functions $\sim Beta(a, \beta)$ given $RDL$, where $a$ is the number of shots and $\beta$ is the number of minutes minus the number of shots.

iv. The node $S$, which represents the expected number of shots, is a Binomial probability mass function $\sim B(n, p)$,

$$f(k, n, p) = Pr(k|n, p) = Pr(S = k) = \left( \frac{n!}{k!\,(n-k)!} \right) p^k (1-p)^{n-k}$$

where $n$ represents the number of minutes in possession of the ball defined as[§] P×90, under the assumption a match lasts 90 playable minutes, and $p$ is $p(SM)$; i.e., the probability to generate a shot per minute spent in possession of the ball, as defined above.

v. The node $p(ST)$, which represents the probability for a shot to be on target, is also a mixture of probability density functions $\sim Beta(a, \beta)$ given $RDL$, where $a$ is the number of shots on target, and $\beta$ is the number of shots off target; i.e., total shots minus shots on target.

vi. The node $ST$, which represents the expected number of shots on target, is also a Binomial probability mass function $\sim B(n, p)$, where $n$ is the expected number of shots $S$ and $p$ is the probability for a shot to be on target $p(ST)$.

vii. The node $p(G)$, which represents the probability to score a goal, is also a mixture of probability density functions $\sim Beta(a, \beta)$ given $RDL$, where $a$ is the number of goals scored, and $\beta$ is the number of shots on target successfully defended; i.e., total shots on target minus goals scored.

viii. The node $G$, which represents the expected number of goals scored, is also a Binomial probability mass function $\sim B(n, p)$, where $n$ is the expected number of shots on target $ST$, and $p$ is the probability to score a goal $p(G)$.

[§] For the away team (i.e., *AT*) it is (1−P)×90.

ix. The node 1X2 is a discrete distribution with states *Home win*, *Draw*, and *Away win*, determined by the distributions $G$ of both home ($H$) and away ($A$) teams; i.e., 1X2="*Home win*" if $G_{oH} > G_{oA}$, "*Away win*" if $G_{oH} < G_{oA}$, "*Draw*" otherwise.

x. The node $GD$, which represents goal difference, is simply $G_{oH} - G_{oA}$.

xi. The node $AH$ represents a set of nodes corresponding to all the possible AH outcomes with state probabilities for home and away wins, given $GD$, as defined in Section 2.

It should be clear by this point that for both home and away teams: a) nodes $P$ and $p(SM)$ are hyperparameters of Beta node $S$, b) nodes $S$ and $p(ST)$ are hyperparameters of Beta node $ST$, and c) nodes $ST$ and $p(G)$ are hyperparameters of node $G$; effectively creating a Beta-Binomial Hybrid BN process.

## 4. Data and model fitting

### *4.1. Data*

The rating method, the BN model, and the betting simulation are based on data collected from www.football-data.co.uk and manually recorded from www.nowgoal.com. Table 5 specifies which of the data variables are used by the rating system, the BN model, and the betting simulation. For example, the rating system only requires information about goal data and hence, it only considers variables *Date*, *Home team*, *Away team*, *Home goals*, and *Away goals*. Since the ratings are used as an input into the BN model, they represent an additional BN variable and at the same time make the BN model independent of variables *Date*, *Home team* and *Away team*.

The data are based on the English Premier League (EPL) seasons 1992/93 to 2018/19. However, AH odds data were available only from season 2006/07 onwards, whereas ball possession data which is needed by the BN model were available only from season 2009/10 onwards. As a result, the rating system is trained with up to 27 seasons of data, the BN is parameterised with up to 10 seasons of data (since it requires possession data), and the betting simulation is performed over 13 seasons (since it requires AH odds data).

**Table 5.** The data variables used to train the rating system (R), the BN model (BN), and to simulate betting (B).

| Variable | Details | Used in |
|---:|:---:|:---:|
| Date | The date of the match | R |
| Home team | The team playing at home grounds | R |
| Away team | The team playing at away grounds | R |
| Match outcome (1X2) | The outcome of the match in terms of home win, draw, or away win | BN, B |
| Home possession rate | The ball possession rate of the team playing at home | BN |
| Away possession rate | The ball possession rate of the team playing away | BN |
| Home shots | The number of shots created by the home team | BN |
| Away shots | The number of shots created by the away team | BN |
| Home shots on target | The number of shots on target created by the home team | BN |
| Away shots on target | The number of shots on target created by the away team | BN |
| Home goals | The number of goals scored by the home team | R, BN, B |
| Away goals | The number of goals scored by the away team | R, BN, B |
| Team ratings | The difference between home team and away team ratings | BN |
| Handicap | The AH on which the market odds are based | B |
| 1X2 odds | The average and maximum (i.e., best available) bookmaker 1X2 odds | B |
| AH odds | The average and maximum (i.e., best available) bookmaker AH odds | B |

### *4.2. Model fitting*

By definition, the ratings are developed in a temporal manner. That is, for a match prediction at time $t$ the model considers team ratings at time $t-1$. For any match prediction, a team's rating will always be based on the most recent rating prior to the date of the match under prediction, and a team's rating will always be based on past match results.

Conversely, the BN model functions as a machine learning model independent of time and is validated using leave-one-out cross validation (LOOCV). A prediction between teams that have rating difference $Z$, where $Z$ is one of the 23 $RDL$s as defined in Table 4, is derived from all data matches with rating difference $Z$, excluding the match under prediction during validation.

This combination of model parameterisation and validation with a rating system and a BN model is adopted by (Constantinou, 2018). The validation approach is unconventional because the BN model assumes no temporal relationships. When applied to past matches, it generates predictions at time $t$ based on the whole data set which may include future match results. The reason this approach works well, without overestimating the future accuracy of the model, is because it does not matter whether the data comes from past or future. This is because the model assumes that the relationship between, for example, shots on target and goals scored remains invariant over time for the average EPL team, and empirical results support this claim. These include:

i. The results presented in Sections 5.2 and 5.3 which show that predictive accuracy is consistent across all 13 seasons tested, including the three seasons 2006/07 to 2008/09 which did *not* form part of the BN's training data;

ii. The model in (Constantinou, 2018) which was based on this approach and ranked 2nd in the international *Machine Learning for Soccer* competition, with a prediction error consistent with the validation error.

The empirical support for the model extends to demonstrating that the model can produce good predictions for matches between teams $X$ and $Y$ even when the prediction is derived from match data that neither $X$ nor $Y$ participated in (Constantinou, 2018). This claim is also supported by the results presented in this paper. Specifically, during Seasons 2006/07 to 2008/09 the following teams have had their performance determined by data that did not include any of their matches: a) Sheffield United, b) Charlton, and c) Derby. The reason this occurred is because, as discussed above, the BN model was trained with data from season 2009/10 onwards, which does not include any match instances associated with these teams. Their performance in terms of possession, shots, shots on target and goals scored was derived by other similar match situations in terms of rating difference between home and away teams.

This approach has advantages and disadvantages. The disadvantage is that, for those who are interested in how such a model would have performed in the past, the results only approximate past performance under the assumption the model would have been trained with at least the same amount of data as the test model. On the other hand, the advantage is that this approach allows us to preserve the sample size of the training data throughout validation, and this enables us to validate how the resulting model would have performed over multiple seasons without modifying its parameterisation (excluding the removal of a single sample; i.e. the match under assessment during validation).

To understand why this is important, consider evaluating match instances five seasons in the past. A temporal model would require the removal of the five most recent seasons from the training data. This would have led to limited samples for some of the predetermined levels of rating difference shown in Table 4. The limited data issue can only be overcome by reducing the number of predetermined levels of rating difference (i.e., the dimensionality of the model); but doing so would produce a different model than the one described. Instead, the approach adopted by (Constantinou, 2018) enables us to address the temporal aspect of the problem through the ratings and to preserve the fitting of the BN across all seasons tested; effectively enabling us to test the current parameterised model on multiple seasons independent of time.

## 5. Results

The results are reported in terms of rating (i.e., goal difference) error, predictive accuracy and profitability. Specifically, Section 5.1 assesses the accuracy of the modified pi-ratings in terms of expected goal difference error, Section 5.2 assesses the accuracy of the overall model in predicting both AH and 1X2 outcomes, and Section 5.3 assesses the capability of the model in terms profitability in both 1X2 and AH markets.

### *5.1. Pi-ratings accuracy and overall model fitting*

Fig 2 shows that the optimal λ and γ parameters, that minimise the goal difference error as defined in Section 3, are λ=0.018 and γ=0.7. Note that while the results are based on training data from seasons 1992/93 to 2018/19, the optimisation is restricted to match instances in which both teams had previously played at least 38 matches; a total of 9,073 match instances. This is to ensure that the model is optimised on matches in which both teams have had their ratings developed by at least one football season.

The optimal learning rates are fairly consistent with those reported in (Constantinou & Fenton, 2013) (i.e., λ=0.035 and γ=0.7) on the basis of minimising goal difference error over five EPL seasons, with those reported in (Van Cutsem, 2019) (i.e., where λ=0.06 and γ=0.6) on the basis of minimising mean squared goal difference error over eight EPL seasons, with those reported in (Constantinou, 2018), λ=0.054 and γ=0.79, on the basis of minimising the Rank Probability Score (RPS) error metric (Constantinou & Fenton, 2012) over multiple leagues worldwide, and with those reported in (Hubacek et al., 2018), λ=0.06 and γ=0.5, where pi-ratings had been used in conjunction with Gradient boosted trees parameters to minimise RPS over multiple leagues worldwide.

However, note that the optimal learning rate λ is lower in this study, and this is likely due to the modification that performs more aggressive rating revisions to the first 38 matches of each team, since it is intended to improve the speed of rating convergence. Interestingly, the overall mean goal difference error shown in Fig 2, $e$=1.2283 (or $e^2$=1.509), is considerably lower than those reported in (Constantinou & Fenton, 2013) and (Van Cutsem, 2019), where $e^2$=2.625 and $e^2$=2.66 respectively, and this suggests that the modifications have had a positive impact on the ratings.

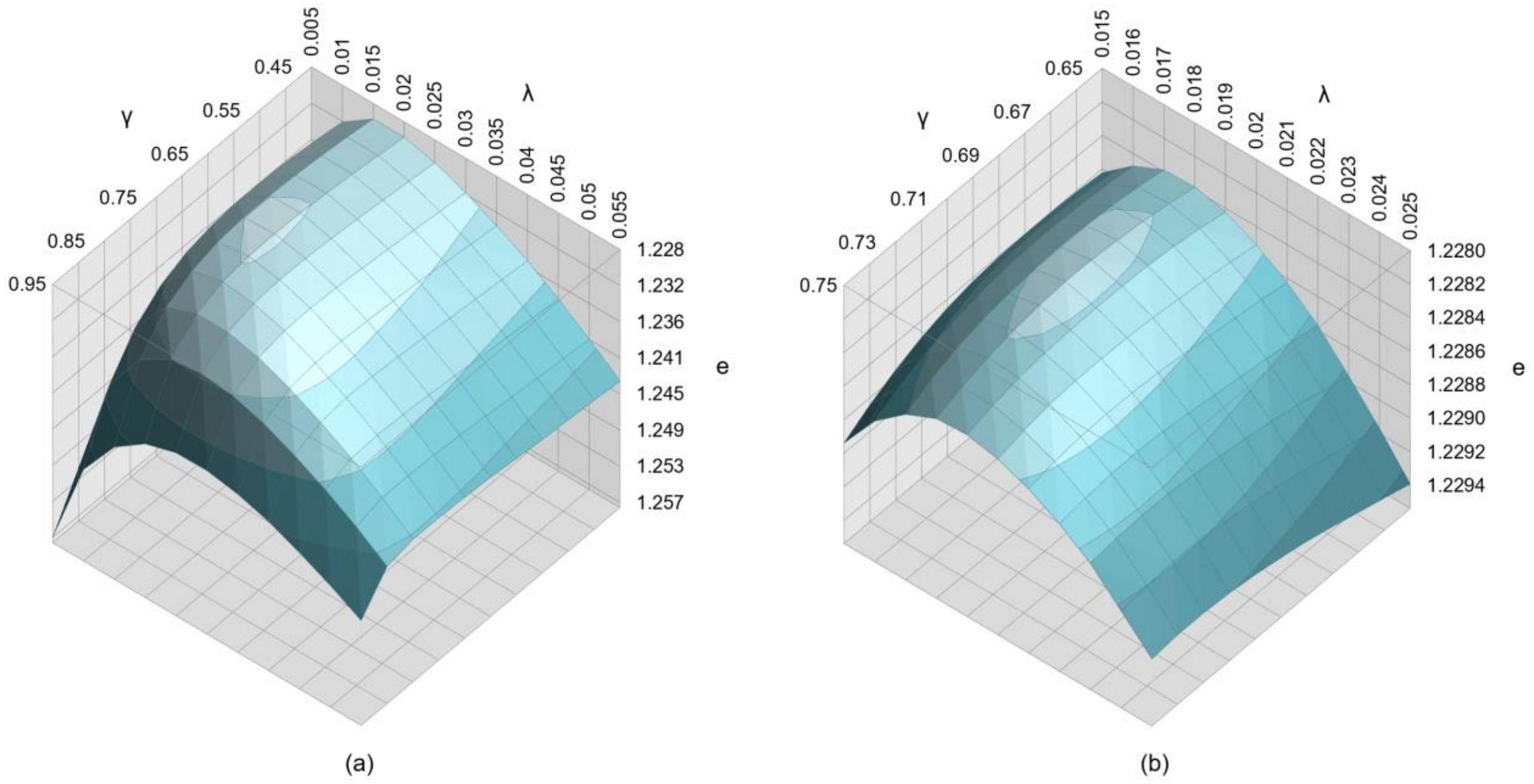

**Figure 2.** The optimal modified pi-rating learning rates and associated prediction error $e$, given k=3, are λ=0.018 and γ=0.7. The results are based on training data from seasons 1992/93 to 2018/19. The optimisation is restricted to match instances in which both teams had previously played at least 38 matches; a total of 9,073 match instances.

Fig 3 illustrates the expected goal difference for each of the 23 rating difference levels. Level 23 represents the highest rating discrepancy in favour of the away team, where the average expectation of the match is a score difference of -1.38 (or 1.38 goals in favour of the away team), and level 1 represents the highest rating discrepancy in favour of the home team with an expected score difference of 2.18 (or 2.18 goals in favour of the home team). The graph reveals a linear relationship between rating discrepancy and score discrepancy.

As shown in Table 4, the granularity of the 23 intervals was selected to ensure that for any rating difference state there are at least 50 data points for a reasonably well-informed prior of observed goal difference. As with any discretised variable, different splits produce slightly different results. In the case of the $RDL$ distribution, any changes in discretisation will remain faithful to the linear relationship illustrated in Fig 3; implying that we should expect minor changes to the interval averages as long as the number of splits remains invariant and data points for each interval are maintained above 50.

Any minor model amendment is naturally expected to have minor impact on the predicted probabilities, and any minor impact is expected to have some influence on the results based on small discrepancies between predicted and published market odds (i.e., small $\theta$ values as defined later in subsection 5.3, such as $\theta = 1$ or 2). However, no changes are expected for larger discrepancies. Since the conclusions in this paper are not driven by results that are based on such small differences between predicted and observed odds, any minor modification is not expected to alter concluding remarks.

Table 6 presents the discrepancy in prediction error $e$ for the different hyperparameter combinations of λ and γ, relative to the prediction error generated by the optimal inputs of λ=0.018 and γ=0.7. Recommended values for λ and γ, that could be considered as an alternative to the optimal values of λ=0.018 and γ=0.7, are the ones that generate up to 0.01% discrepancy in prediction error. A total of 31 different combinations, excluding the optimal combination, generate discrepancy within the 0.01% threshold. By keeping the value for λ static, the recommended hyperparameter combinations are a) λ=0.017 and γ=0.65−0.73, b) λ=0.018 and γ=0.65−0.74, c) λ=0.019 and γ=0.65−0.74, and d) λ=0.02 and γ=0.7−0.72.

**Table 6.** The discrepancy in prediction error $e$ for the different combinations of λ and γ, relative to the optimal inputs of λ=0.018 and γ=0.7. Darker green cells represent lower discrepancy, whereas darker red cells represent higher discrepancy. Recommended inputs for λ and γ are the ones that generate up to 0.01% discrepancy.

| λ \ γ | **0.65** | **0.66** | **0.67** | **0.68** | **0.69** | **0.7** | **0.71** | **0.72** | **0.73** | **0.74** | **0.75** |
|---|---|---|---|---|---|---|---|---|---|---|---|
| **0.015** | 0.026% | 0.025% | 0.024% | 0.024% | 0.025% | 0.026% | 0.028% | 0.032% | 0.037% | 0.043% | 0.051% |
| **0.016** | 0.014% | 0.013% | 0.012% | 0.011% | 0.011% | 0.012% | 0.014% | 0.016% | 0.020% | 0.025% | 0.032% |
| **0.017** | 0.007% | 0.005% | 0.004% | 0.003% | 0.003% | 0.003% | 0.004% | 0.006% | 0.009% | 0.014% | 0.020% |
| **0.018** | 0.007% | 0.004% | 0.002% | 0.001% | 0.000% | 0.000% | 0.001% | 0.003% | 0.005% | 0.009% | 0.013% |
| **0.019** | 0.010% | 0.007% | 0.006% | 0.004% | 0.004% | 0.004% | 0.004% | 0.005% | 0.006% | 0.009% | 0.012% |
| **0.020** | 0.017% | 0.015% | 0.013% | 0.012% | 0.010% | 0.009% | 0.009% | 0.009% | 0.010% | 0.012% | 0.015% |
| **0.021** | 0.026% | 0.024% | 0.022% | 0.020% | 0.019% | 0.017% | 0.017% | 0.016% | 0.017% | 0.018% | 0.021% |
| **0.022** | 0.038% | 0.035% | 0.033% | 0.031% | 0.029% | 0.028% | 0.026% | 0.025% | 0.026% | 0.027% | 0.029% |
| **0.023** | 0.053% | 0.050% | 0.046% | 0.044% | 0.041% | 0.040% | 0.038% | 0.037% | 0.037% | 0.037% | 0.039% |
| **0.024** | 0.069% | 0.066% | 0.062% | 0.059% | 0.056% | 0.054% | 0.052% | 0.050% | 0.050% | 0.051% | 0.053% |
| **0.025** | 0.087% | 0.083% | 0.080% | 0.076% | 0.073% | 0.070% | 0.068% | 0.066% | 0.066% | 0.066% | 0.067% |

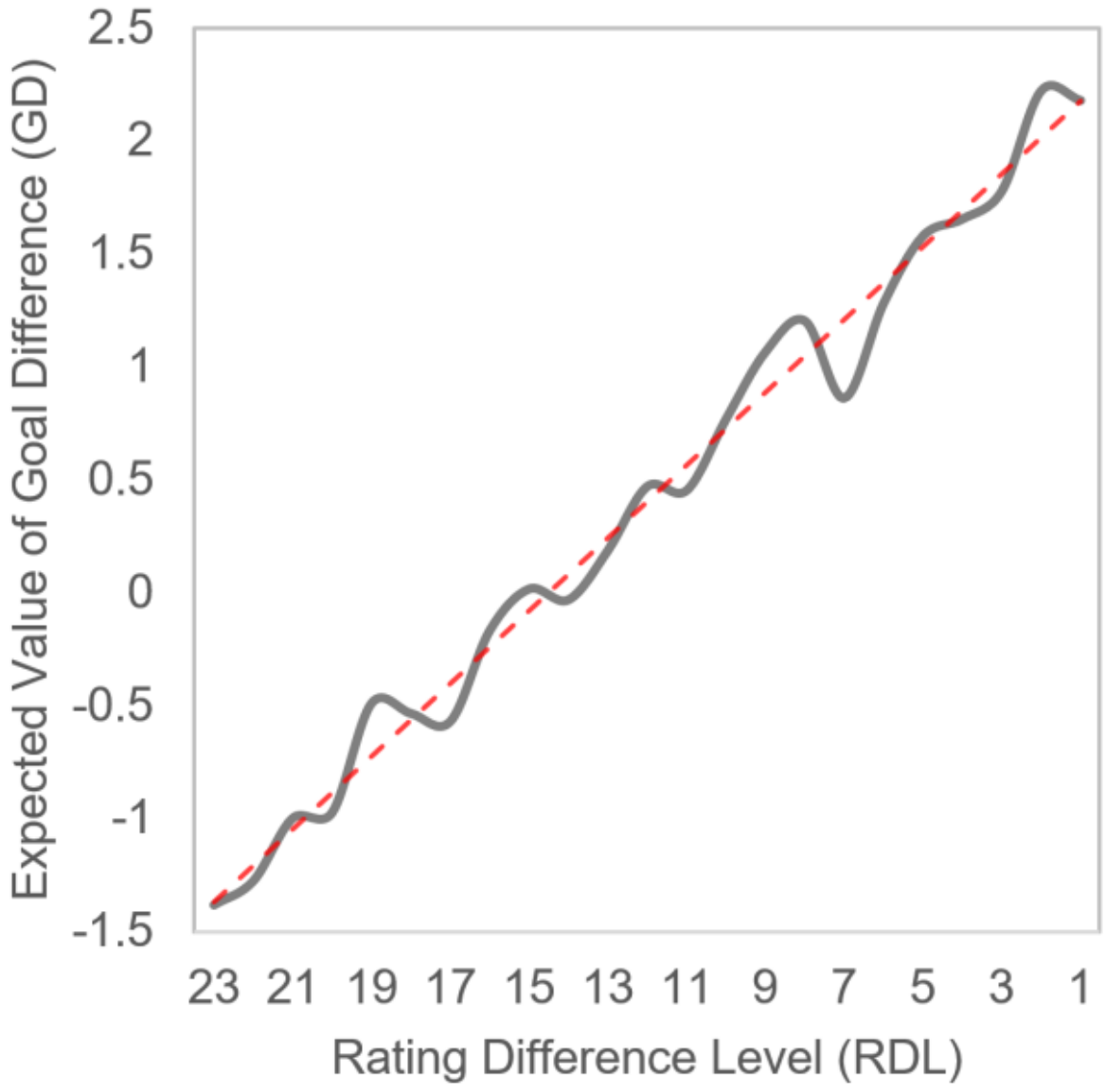

**Figure 3.** Sensitivity analysis between the 23 states of $RDL$ node and the expected goal difference, with linear trendline superimposed as a dashed red line.

### *5.2. Predictive accuracy*

Predictive accuracy is measured for both 1X2 and AH distributions. The Brier Score is used to measure the accuracy of the binary AH outcome, and the RPS metric (Constantinou & Fenton, 2012) is used to measure the accuracy of the multinomial 1X2 distribution. The RPS can be viewed as a Brier Score extended to multinomial ordinal distributions.

Table 7 shows that the RPS error for the 1X2 outcomes ranges from 0.184 to 0.213 with an average RPS of 0.195 across all 13 EPL seasons. This result compares well relative to previous studies that assumed pi-ratings. Specifically, in (Constantinou, 2018) the RPS ranged from 0.187 to 0.236 for 52 different leagues worldwide, with an average RPS of 0.211 at validation, an average RPS of 0.203 for EPL matches, and an average RPS of 0.208 in the competition. Similarly, the average RPS was ~0.2 in (Hubacek et al., 2018), according to Fig 3, and 0.206 in the competition.

**Table 7.** Predictive accuracy across all seasons, based on the Rank Probability Score (RPS) for multinomial 1X2 predictions and the Brier Score (BS) for binary AH predictions. Lower score indicates higher predictive accuracy for both RPS and BS.

| Season | RPS (1X2 accuracy) | BS (AH accuracy) |
|---|---|---|
| 2006/07 | 0.197 | 0.252 |
| 2007/08 | 0.184 | 0.248 |
| 2008/09 | 0.192 | 0.229 |
| 2009/10 | 0.188 | 0.199 |
| 2010/11 | 0.202 | 0.248 |
| 2011/12 | 0.205 | 0.257 |
| 2012/13 | 0.191 | 0.258 |
| 2013/14 | 0.195 | 0.249 |
| 2014/15 | 0.199 | 0.254 |
| 2015/16 | 0.213 | 0.254 |
| 2016/17 | 0.191 | 0.267 |
| 2017/18 | 0.192 | 0.253 |
| 2018/19 | 0.191 | 0.260 |
| **Overall** | **0.195** | **0.248** |

These results are consistent with those reported in Section 5.1, which show that the overall error $e$ optimises lower in this study; i.e., the ratings more accurately predict score difference. The results from profitability presented in Section 5.3 are also consistent with these findings.

### *5.2.1 Time-series analysis and sample size requirements*

The model is trained with data the covers 13 years of data, and not all the variables could be measured throughout this period. Still, the model seems to work well without evidence of bias. This subsection investigates whether the variables used in the model show any drift over time that the model might have filtered out. Moreover, because the dimensionality of the model has been adjusted relative to the available sample size, this subsection also reports the sample size required for the model priors to be well informed by data.

Fig 4 presents the results from time-series analysis in investigating potential shifts in the predictive outputs of the model over time. The analysis is performed by increasing the training data set by a single football season's worth of data at a time, and the shift is measured in terms of changes in the expected value of the given distribution. As shown in Fig 4, the analysis focuses on the four main variables of the BN model; namely possession, shots, shots on target, and goals scored. Moreover, the different levels of rating difference (refer to Table 4) are categorised into four groups, and are measured with reference to the 10 specified seasons.

The reason the first three, out of the 13, seasons are not considered here is because (as later shown in Table 8) the BN model was not trained with data samples from the first two seasons, and this also means that any results obtained during the third season rely on very low samples. As previously discussed in subsection 3.2, the reason the first two seasons are not considered by the BN model is because the BN is trained with match instances in which both teams had previously played at least 38 matches, to allow for the pi-ratings to converge to reasonably accurate estimates before considered for model training.

The results are discussed with reference to Table 8, which presents the available samples for each of the 23 levels of rating difference, after each subsequent league data set is added to the training data set that was used to learn the BN model. The results show that many of the shifts occur in the first few seasons, and this is reasonable since the first seasons are the ones which rely on fewer samples. Shifts are also observed after adding the most recent leagues, but these shifts are largely restricted to the outputs of possession and shots on target, and involve matches with a level of rating difference between 1 to 6. The most important output of the model, which is the goal difference, remains essentially unchanged over time. Therefore, while it is reasonable to assume that shifts in playing style might occur naturally from season to season, collectively these shifts appear to have no meaningful impact on the prediction of goal difference, from which the 1X2 and AH distributions are formed.

Lastly, the results in Table 8 suggest that the model described in this paper should be trained with at least five seasons of league data, to ensure that the model priors are well informed by data. Because the model is trained with publicly available data, and which only increases over time, there is no motivation to use less data than what is currently available and hence, this requirement should not be viewed as a limitation.

**Table 8.** The number of data samples observed in each of the 23 intervals of Rating Difference Level (RDL) after each subsequent league data set is added to the training data. Darker red cells represent sample size less than 19 (equivalent to less than half a season's data), orange cells represent sample size less than 38 (equivalent to more than half, and less than, a season's data), and white cells represent sufficient sample size.

| Season/ | Rating Difference Level (RDL) | | | | | | | | | | | | | | | | | | | | | | |
|---|---|---|---|---|---|---|---|---|---|---|---|---|---|---|---|---|---|---|---|---|---|---|---|
| **Data set** | **1** | **2** | **3** | **4** | **5** | **6** | **7** | **8** | **9** | **10** | **11** | **12** | **13** | **14** | **15** | **16** | **17** | **18** | **19** | **20** | **21** | **22** | **23** |
| 2006/07 | 0 | 0 | 0 | 0 | 0 | 0 | 0 | 0 | 0 | 0 | 0 | 0 | 0 | 0 | 0 | 0 | 0 | 0 | 0 | 0 | 0 | 0 | 0 |
| 2007/08 | 0 | 0 | 0 | 0 | 0 | 0 | 0 | 0 | 0 | 0 | 0 | 0 | 0 | 0 | 0 | 0 | 0 | 0 | 0 | 0 | 0 | 0 | 0 |
| 2008/09 | 3 | 9 | 17 | 10 | 9 | 11 | 12 | 9 | 19 | 16 | 31 | 28 | 31 | 20 | 30 | 17 | 13 | 9 | 14 | 8 | 9 | 9 | 8 |
| 2009/10 | 14 | 19 | 24 | 20 | 18 | 20 | 33 | 22 | 33 | 38 | 69 | 50 | 51 | 48 | 60 | 34 | 26 | 19 | 24 | 12 | 18 | 16 | 16 |
| 2010/11 | 20 | 25 | 36 | 27 | 40 | 30 | 44 | 35 | 54 | 65 | 88 | 78 | 81 | 74 | 82 | 51 | 45 | 28 | 32 | 21 | 27 | 23 | 20 |
| 2011/12 | 24 | 32 | 46 | 35 | 58 | 45 | 63 | 48 | 72 | 89 | 120 | 115 | 120 | 104 | 103 | 68 | 59 | 41 | 47 | 28 | 41 | 25 | 23 |
| 2012/13 | 29 | 37 | 62 | 46 | 71 | 62 | 72 | 61 | 83 | 109 | 151 | 152 | 152 | 117 | 128 | 82 | 72 | 51 | 65 | 43 | 48 | 28 | 27 |
| 2013/14 | 31 | 46 | 71 | 64 | 86 | 74 | 86 | 80 | 99 | 132 | 175 | 178 | 187 | 151 | 158 | 96 | 93 | 60 | 78 | 54 | 60 | 35 | 34 |
| 2014/15 | 35 | 50 | 76 | 73 | 96 | 90 | 101 | 103 | 121 | 149 | 201 | 208 | 211 | 180 | 181 | 114 | 113 | 75 | 90 | 64 | 67 | 37 | 35 |
| 2015/16 | 35 | 51 | 86 | 82 | 115 | 111 | 114 | 116 | 139 | 176 | 218 | 244 | 251 | 216 | 204 | 135 | 130 | 90 | 104 | 84 | 74 | 40 | 35 |
| 2016/17 | 39 | 58 | 97 | 94 | 130 | 122 | 124 | 126 | 150 | 189 | 238 | 275 | 277 | 248 | 220 | 147 | 138 | 96 | 120 | 100 | 81 | 50 | 37 |
| 2017/18 | 55 | 68 | 107 | 107 | 144 | 138 | 134 | 135 | 162 | 208 | 263 | 306 | 309 | 278 | 245 | 162 | 153 | 110 | 131 | 115 | 93 | 58 | 55 |

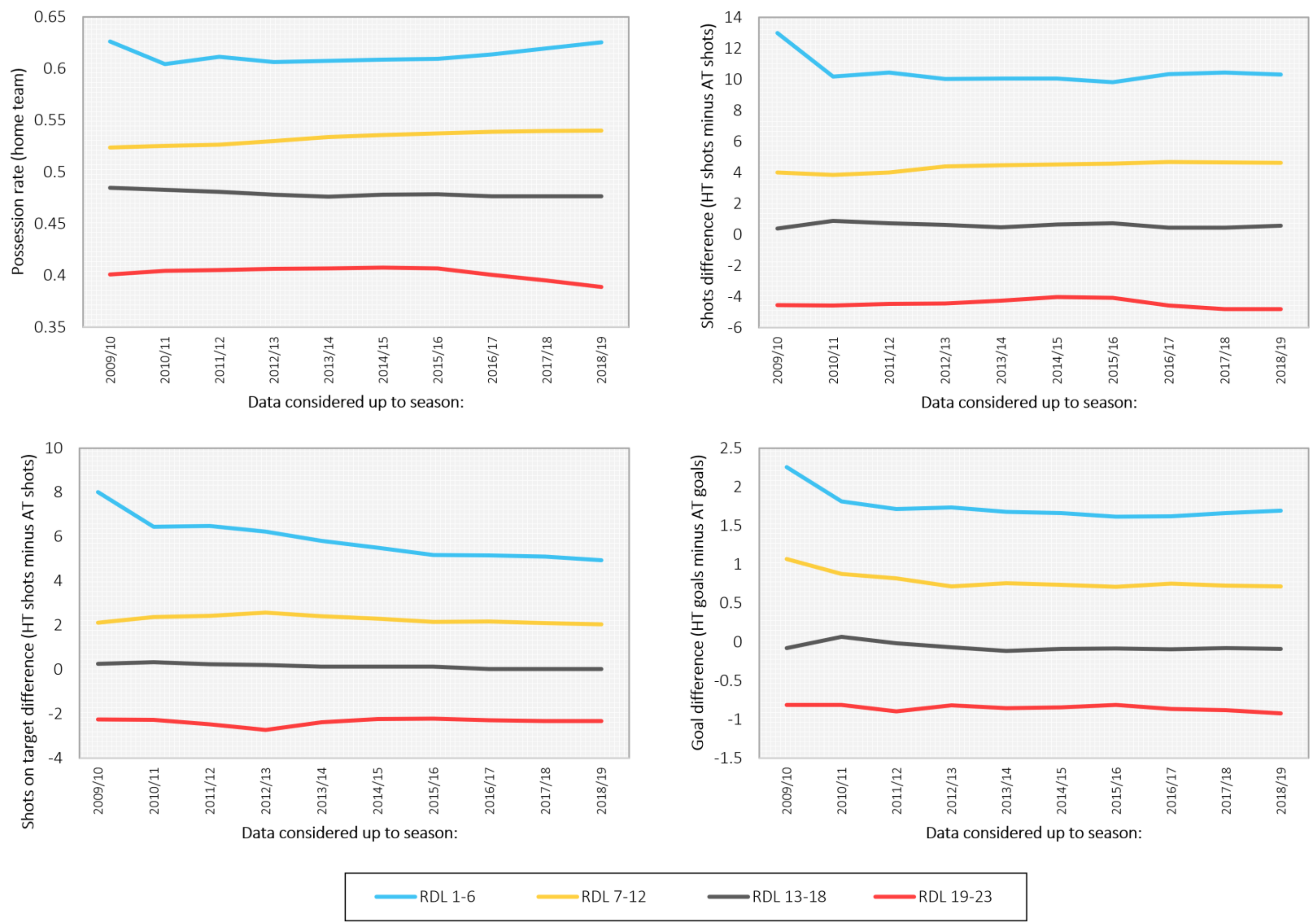

**Figure 4.** Investigating the prediction shift over time, with reference to the four main variables of the BN model. The shift is measured in terms of the expected value of the specified distribution, and by adding a football season's worth of data, to the training data set, at a time.

### *5.3. Profitability*

The assessment of profitability is based on 13 EPL seasons and considers both the average and the best available (maximum) market odds. The simulation is evaluated both in terms of maximising profit as well as the Return On Investment (ROI). A standard betting strategy is used where fixed singe-unit bets (e.g., £1) are placed on 1X2 and AH outcomes with payoff that is higher than the model's estimated unbiased payoff by at least $\theta$, where $\theta$ is the discrepancy between the predicted probability and the payoff probability. For example, if the model predicts 51% and the bookmakers' payoff for that event is 50% (i.e., odds of 2), then $\theta = 1$; i.e., the bookmakers pay 1% more than the model's estimate. The betting simulation is performed across all payoff decision thresholds $\theta$, for both 1X2 and AH outcomes. The results are first discussed in terms of 1X2 betting performance with reference to Tables 9, 10 and 11; then in terms of AH betting performance with reference to Tables 12, 13 and 14.

Table 9 presents the profitability generated by 1X2 bets over all possible payoff decision thresholds $\theta$, assuming static $\theta$ across all 13 seasons, for both average and maximum market odds. Unsurprisingly, the results show that it is much easier for the model to generate profit when taking advantage of the maximum market odds. Moreover, low thresholds $\theta$ (i.e., when the predictions are roughly in agreement with market odds) are not profitable. Interestingly, ROI maximises at much higher thresholds $\theta$ compared to profit; i.e., profit maximises at 8% and 9% whereas ROI at 18% and 16%, for average and maximum market odds respectively. This is because lower thresholds $\theta$ generate a higher number of bets which can generate higher profit even if ROI is lower.

Tables 10 and 11 show how the profitability changes when we consider the threshold $\theta$ that maximises ROI (Table 10) or profit (Table 11) per season, rather than considering a static $\theta$ across all seasons (Table 9), for both average and maximum market odds. Profit, once more, tends to maximise on lower thresholds $\theta$ compared to ROI. As an example, Table A1 provides the information used during the betting simulation to assess profitability for 1X2 outcomes, based on average odds of season 2010/11 as shown in Table 10.

The results show that the optimal threshold $\theta$ varies dramatically between seasons, and there is much to be gained by identifying the optimal $\theta$. However, the high variance of $\theta$ suggests that is *not* reasonable to expect that we will be able to successfully predict the optimal value of $\theta$ before a season starts. Moreover, while the results are restricted to cases with 30 or more bets in a single season, it is clear that in many cases the sample size remains insufficient for deriving reliable and robust single-season conclusions. This means that the maximised profitability presented in Tables 10 and 11 is not a realistic expectation of real-world performance; only Table 9 is. These results are important because they highlight the danger when optimising models based on the results of a single season (or few seasons), which is often the case in the literature. This outcome is also discussed in Section 6, point *iv*.

Interestingly, while maximising profit per season is guaranteed to also maximise profit over all seasons (Table 11), the same does not apply to ROI (see Table 10 and compare it to Table 11). That is, optimising the betting strategy for maximum ROI per season does not necessarily imply that ROI will maximise across all seasons. Table 10 shows that even though ROI is maximised for each season independently, the overall ROI across all 13 seasons is 9.03% in the case of average odds, which is notably lower compared to the respective overall ROI of 12.96% in Table 11. However, this observation does not extend to the case of maximum market odds. This outcome is also discussed in Section 6, point *vi*.

Tables 12, 13, and 14 repeat the analysis of Tables 9, 10, and 11, but for AH rather than 1X2 betting. Table A2 presents an example of the information used during the betting simulation to assess profitability from AH bets, and it is based on average odds of season 2010/11 as shown in Table 13. Overall, the AH bets appear to generate both lower profit as well as ROI compared to 1X2 bets. As with 1X2 bets, optimising for maximum ROI per season leads to a lower ROI across all seasons, compared to maximising profit. Specifically, Table 13 shows that when maximising ROI per season leads to an overall ROI of 5.7% for average odds, which is lower than the respective average ROI of 6.33% in Table 14 when maximising profit. Once more, this observation only applies to the case of average market odds.

An interesting observation is that AH betting generates a lower number of bets when $\theta$ is low, compared to 1X2 betting, and a higher number of bets when $\theta$ is high. This suggests that AH betting is less sensitive to the betting decision threshold $\theta$ compared to 1X2 betting, for both average and maximum market odds. Furthermore, the optimal threshold $\theta$ for AH bets does not vary as much as it did for 1X2 bets. Despite the relatively low variance of $\theta$ and the common occurrence of winning 60+ out of 100 AH bets, profitability is still inconsistent between seasons. This is because match bets do not share the same payoff.

**Table 9.** Profitability based on average (left) and maximum (right) market odds for **1X2 bets** simulated over 13 EPL seasons; from 2006/09 to 2018/19.

| $\theta$ | Average market odds | | | | | | | Maximum market odds | | | | | | |
|---|---|---|---|---|---|---|---|---|---|---|---|---|---|---|
| | Bets | Bets won | Odds | Win Rate | Returns | Profit | ROI | Bets | Bets won | Odds | Win Rate | Returns | Profit | ROI |
| 0% | 4334 | 1290 | 3.06 | 29.8% | 3942.5 | -391.5 | -9.03% | 4938 | 1514 | 3.22 | 30.7% | 4879.0 | -59.01 | -1.20% |
| 1% | 3712 | 1105 | 3.06 | 29.8% | 3375.9 | -336.1 | -9.05% | 4794 | 1468 | 3.24 | 30.6% | 4755.3 | -38.67 | -0.81% |
| 2% | 3109 | 937 | 3.11 | 30.1% | 2909.9 | -199.2 | -6.41% | 4305 | 1307 | 3.19 | 30.4% | 4167.1 | -137.91 | -3.20% |
| 3% | 2538 | 775 | 3.09 | 30.5% | 2397.8 | -140.2 | -5.52% | 3658 | 1122 | 3.23 | 30.7% | 3623.2 | -34.81 | -0.95% |
| 4% | 2072 | 643 | 3.09 | 31.0% | 1987.2 | -84.9 | -4.10% | 3067 | 950 | 3.26 | 31.0% | 3096.4 | 29.38 | 0.96% |
| 5% | 1699 | 524 | 3.15 | 30.8% | 1650.9 | -48.1 | -2.83% | 2553 | 802 | 3.27 | 31.4% | 2620.0 | 67.02 | 2.63% |
| 6% | 1339 | 415 | 3.19 | 31.0% | 1325.3 | -13.7 | -1.02% | 2076 | 656 | 3.29 | 31.6% | 2157.8 | 81.77 | 3.94% |
| 7% | 1036 | 329 | 3.26 | 31.8% | 1074.0 | 38.0 | 3.67% | 1682 | 526 | 3.34 | 31.3% | 1756.3 | 74.29 | 4.42% |
| 8% | 814 | 260 | 3.30 | 31.9% | 858.7 | 44.7 | 5.49% | 1345 | 422 | 3.42 | 31.4% | 1444.6 | 99.56 | 7.40% |
| 9% | 612 | 198 | 3.17 | 32.4% | 626.8 | 14.8 | 2.42% | 1049 | 352 | 3.48 | 33.6% | 1226.2 | 177.18 | 16.89% |
| 10% | 452 | 148 | 3.13 | 32.7% | 462.6 | 10.6 | 2.34% | 807 | 267 | 3.40 | 33.1% | 906.6 | 99.63 | 12.35% |
| 11% | 320 | 101 | 3.12 | 31.6% | 315.2 | -4.8 | -1.51% | 608 | 199 | 3.28 | 32.7% | 652.3 | 44.29 | 7.28% |
| 12% | 241 | 75 | 3.16 | 31.1% | 236.9 | -4.1 | -1.68% | 451 | 155 | 3.30 | 34.4% | 511.8 | 60.78 | 13.48% |
| 13% | 191 | 65 | 3.27 | 34.0% | 212.3 | 21.3 | 11.13% | 324 | 108 | 3.27 | 33.3% | 352.7 | 28.65 | 8.84% |
| 14% | 143 | 50 | 3.45 | 35.0% | 172.5 | 29.5 | 20.59% | 241 | 80 | 3.38 | 33.2% | 270.4 | 29.37 | 12.19% |
| 15% | 103 | 37 | 3.10 | 35.9% | 114.8 | 11.8 | 11.48% | 184 | 62 | 3.48 | 33.7% | 216.0 | 31.98 | 17.38% |
| 16% | 76 | 27 | 3.40 | 35.5% | 91.8 | 15.8 | 20.84% | 132 | 46 | 3.48 | 34.8% | 160.0 | 27.99 | 21.20% |
| 17% | 51 | 17 | 3.64 | 33.3% | 61.8 | 10.8 | 21.20% | 101 | 37 | 3.31 | 36.6% | 122.3 | 21.31 | 21.10% |
| 18% | 37 | 12 | 3.81 | 32.4% | 45.7 | 8.7 | 23.59% | 76 | 26 | 3.46 | 34.2% | 89.9 | 13.91 | 18.30% |
| 19% | 24 | 7 | 3.10 | 29.2% | 21.7 | -2.3 | -9.58% | 52 | 18 | 3.82 | 34.6% | 68.7 | 16.67 | 32.06% |
| 20% | 19 | 5 | 3.44 | 26.3% | 17.2 | -1.8 | -9.37% | 34 | 11 | 3.51 | 32.4% | 38.6 | 4.61 | 13.56% |

**Table 10.** The payoff discrepancies $\theta$ that maximise **ROI** per season (in yellow shading), based on **1X2 bets** and for both average (left) and maximum (right) market odds. The optimal $\theta$ discrepancy is chosen over all $\theta$ that generate at least 30 bets in a single season.

| Season | Average market odds | | | | | | | | Maximum market odds | | | | | | | |
|---|---|---|---|---|---|---|---|---|---|---|---|---|---|---|---|---|
| | $\theta$ | Bets | Bets won | Odds | Win Rate | Returns | Profit | ROI | $\theta$ | Bets | Bets won | Odds | Win Rate | Returns | Profit | ROI |
| 2006/07 | 8% | 34 | 13 | 3.4 | 38.2% | 43.7 | 9.66 | 28.41% | 10% | 39 | 16 | 3.4 | 41.0% | 54.6 | 15.6 | 40.00% |
| 2007/08 | 7% | 63 | 21 | 2.8 | 33.3% | 58.4 | -4.56 | -7.24% | 11% | 41 | 15 | 3.1 | 36.6% | 47.1 | 6.11 | 14.90% |
| 2008/09 | 3% | 181 | 72 | 3.1 | 39.8% | 224.3 | 43.3 | 23.92% | 5% | 202 | 77 | 3.5 | 38.1% | 268.0 | 66.04 | 32.69% |
| 2009/10 | 2% | 242 | 57 | 3.5 | 23.6% | 199.9 | -42.09 | -17.39% | 13% | 30 | 9 | 3.7 | 30.0% | 33.0 | 3.02 | 10.07% |
| 2010/11 | 10% | 33 | 17 | 2.8 | 51.5% | 48.2 | 15.18 | 46.00% | 10% | 59 | 29 | 3.0 | 49.2% | 86.6 | 27.56 | 46.71% |
| 2011/12 | 0% | 326 | 103 | 3.3 | 31.6% | 338.7 | 12.68 | 3.89% | 1% | 370 | 119 | 3.6 | 32.2% | 433.4 | 63.4 | 17.14% |
| 2012/13 | 7% | 65 | 22 | 3.5 | 33.8% | 77.9 | 12.89 | 19.83% | 9% | 71 | 25 | 3.8 | 35.2% | 94.5 | 23.45 | 33.03% |
| 2013/14 | 10% | 30 | 10 | 3.7 | 33.3% | 36.6 | 6.59 | 21.97% | 12% | 33 | 13 | 3.8 | 39.4% | 49.7 | 16.7 | 50.61% |
| 2014/15 | 7% | 73 | 33 | 3.1 | 45.2% | 101.6 | 28.58 | 39.15% | 9% | 72 | 33 | 3.3 | 45.8% | 109.5 | 37.51 | 52.10% |
| 2015/16 | 10% | 65 | 25 | 3.0 | 38.5% | 74.7 | 9.65 | 14.85% | 12% | 57 | 21 | 3.2 | 36.8% | 67.7 | 10.71 | 18.79% |
| 2016/17 | 9% | 58 | 14 | 4.8 | 24.1% | 67.7 | 9.72 | 16.76% | 10% | 80 | 19 | 5.0 | 23.8% | 94.7 | 14.73 | 18.41% |
| 2017/18 | 8% | 85 | 24 | 3.9 | 28.2% | 93.3 | 8.28 | 9.74% | 13% | 31 | 11 | 3.5 | 35.5% | 38.4 | 7.43 | 23.97% |
| 2018/19 | 13% | 32 | 11 | 3.5 | 34.4% | 38.3 | 6.28 | 19.63% | 14% | 34 | 11 | 3.7 | 32.4% | 40.4 | 6.36 | 18.71% |
| **Overall** | **4.35%** | **1287** | **422** | **3.38** | **32.79%** | **1403.16** | **116.16** | 9.03% | **8.01%** | **1119** | **398** | **3.59** | **35.57%** | **1417.62** | **298.62** | 26.69% |

**Table 11.** The payoff discrepancies $\theta$ that maximise **profit** per season (in yellow shading), based on **1X2 bets** and for both average (left) and maximum (right) market odds. The optimal $\theta$ discrepancy is chosen over all $\theta$ that generate at least 30 bets in a single season.

| Season | Average bookmaker odds | | | | | | | | Maximum bookmaker odds | | | | | | | |
|---|---|---|---|---|---|---|---|---|---|---|---|---|---|---|---|---|
| | $\theta$ | Bets | Bets won | Odds | Win Rate | Returns | Profit | ROI | $\theta$ | Bets | Bets won | Odds | Win Rate | Returns | Profit | ROI |
| 2006/07 | 8% | 34 | 13 | 3.4 | 38.2% | 43.7 | 9.66 | 28.41% | 5% | 166 | 56 | 3.4 | 33.7% | 188.4 | 22.39 | 13.49% |
| 2007/08 | 8% | 45 | 14 | 2.9 | 31.1% | 40.8 | -4.16 | -9.24% | 11% | 41 | 15 | 3.1 | 36.6% | 47.1 | 6.11 | 14.90% |
| 2008/09 | 3% | 181 | 72 | 3.1 | 39.8% | 224.3 | 43.3 | 23.92% | 5% | 202 | 77 | 3.5 | 38.1% | 268.0 | 66.04 | 32.69% |
| 2009/10 | 10% | 41 | 11 | 3.1 | 26.8% | 33.7 | -7.28 | -17.76% | 13% | 30 | 9 | 3.7 | 30.0% | 33.0 | 3.02 | 10.07% |
| 2010/11 | 8% | 59 | 27 | 2.8 | 45.8% | 75.9 | 16.85 | 28.56% | 6% | 164 | 63 | 3.1 | 38.4% | 197.2 | 33.19 | 20.24% |
| 2011/12 | 0% | 326 | 103 | 3.3 | 31.6% | 338.7 | 12.68 | 3.89% | 1% | 370 | 119 | 3.6 | 32.2% | 433.4 | 63.4 | 17.14% |
| 2012/13 | 7% | 65 | 22 | 3.5 | 33.8% | 77.9 | 12.89 | 19.83% | 9% | 71 | 25 | 3.8 | 35.2% | 94.5 | 23.45 | 33.03% |
| 2013/14 | 5% | 123 | 44 | 3.3 | 35.8% | 143.2 | 20.21 | 16.43% | 0% | 380 | 127 | 3.4 | 33.4% | 438.1 | 58.11 | 15.29% |
| 2014/15 | 7% | 73 | 33 | 3.1 | 45.2% | 101.6 | 28.58 | 39.15% | 8% | 91 | 40 | 3.3 | 44.0% | 133.2 | 42.22 | 46.40% |
| 2015/16 | 8% | 94 | 34 | 3.1 | 36.2% | 104.7 | 10.66 | 11.34% | 1% | 371 | 130 | 3.1 | 35.0% | 400.4 | 29.41 | 7.93% |
| 2016/17 | 9% | 58 | 14 | 4.8 | 24.1% | 67.7 | 9.72 | 16.76% | 10% | 80 | 19 | 5.0 | 23.8% | 94.7 | 14.73 | 18.41% |
| 2017/18 | 8% | 85 | 24 | 3.9 | 28.2% | 93.3 | 8.28 | 9.74% | 9% | 93 | 26 | 4.2 | 28.0% | 108.0 | 15.04 | 16.17% |
| 2018/19 | 4% | 188 | 67 | 3.1 | 35.6% | 204.5 | 16.46 | 8.76% | 5% | 206 | 73 | 3.0 | 35.4% | 222.3 | 16.25 | 7.89% |
| **Overall** | **4.62%** | **1372** | **478** | **3.28** | **34.84%** | **1549.85** | 177.85 | **12.96%** | **4.61%** | **2265** | **779** | **3.52** | **34.39%** | **2658.36** | 393.36 | **17.37%** |

**Table 12.** Profitability based on average (left) and maximum (right) market odds for **AH bets** simulated over 13 EPL seasons; from 2006/07 to 2018/19.

| | Average market odds | | | | | | | Maximum market odds | | | | | | |
|---|---|---|---|---|---|---|---|---|---|---|---|---|---|---|
| **θ** | **Bets** | **Bets won** | **Odds** | **Win Rate** | **Returns** | **Profit** | **ROI** | **Bets** | **Bets won** | **Odds** | **Win Rate** | **Returns** | **Profit** | **ROI** |
| 0% | 3914 | 2329 | 1.62 | 59.5% | 3762.8 | -151.22 | -3.86% | 4703 | 2830 | 1.66 | 60.2% | 4707.0 | 4.01 | 0.09% |
| 1% | 3471 | 2051 | 1.62 | 59.1% | 3326.5 | -144.52 | -4.16% | 4228 | 2527 | 1.66 | 59.8% | 4203.2 | -24.76 | -0.59% |
| 2% | 3064 | 1817 | 1.61 | 59.3% | 2929.4 | -134.61 | -4.39% | 3788 | 2247 | 1.66 | 59.3% | 3736.8 | -51.25 | -1.35% |
| 3% | 2665 | 1584 | 1.61 | 59.4% | 2554.4 | -110.65 | -4.15% | 3375 | 1998 | 1.67 | 59.2% | 3340.7 | -34.28 | -1.02% |
| 4% | 2286 | 1357 | 1.62 | 59.4% | 2201.3 | -84.71 | -3.71% | 2974 | 1766 | 1.67 | 59.4% | 2949.3 | -24.73 | -0.83% |
| 5% | 1932 | 1157 | 1.62 | 59.9% | 1873.4 | -58.56 | -3.03% | 2586 | 1526 | 1.67 | 59.0% | 2545.8 | -40.17 | -1.55% |
| 6% | 1640 | 984 | 1.63 | 60.0% | 1600.9 | -39.09 | -2.38% | 2192 | 1304 | 1.67 | 59.5% | 2171.3 | -20.74 | -0.95% |
| 7% | 1386 | 837 | 1.63 | 60.4% | 1368.5 | -17.52 | -1.26% | 1883 | 1124 | 1.67 | 59.7% | 1877.9 | -5.15 | -0.27% |
| 8% | 1135 | 688 | 1.63 | 60.6% | 1124.2 | -10.78 | -0.95% | 1587 | 955 | 1.68 | 60.2% | 1600.0 | 13.03 | 0.82% |
| 9% | 936 | 572 | 1.64 | 61.1% | 938.5 | 2.45 | 0.26% | 1307 | 790 | 1.68 | 60.4% | 1325.1 | 18.06 | 1.38% |
| 10% | 765 | 460 | 1.65 | 60.1% | 759.0 | -5.98 | -0.78% | 1080 | 658 | 1.69 | 60.9% | 1111.7 | 31.73 | 2.94% |
| 11% | 614 | 368 | 1.66 | 59.9% | 612.1 | -1.89 | -0.31% | 896 | 541 | 1.70 | 60.4% | 920.6 | 24.61 | 2.75% |
| 12% | 488 | 292 | 1.65 | 59.8% | 480.9 | -7.12 | -1.46% | 728 | 427 | 1.71 | 58.7% | 728.3 | 0.25 | 0.03% |
| 13% | 394 | 233 | 1.63 | 59.1% | 379.7 | -14.29 | -3.63% | 571 | 330 | 1.71 | 57.8% | 564.5 | -6.46 | -1.13% |
| 14% | 312 | 191 | 1.65 | 61.2% | 314.5 | 2.49 | 0.80% | 463 | 279 | 1.71 | 60.3% | 478.3 | 15.26 | 3.30% |
| 15% | 250 | 155 | 1.65 | 62.0% | 255.1 | 5.07 | 2.03% | 369 | 222 | 1.70 | 60.2% | 376.8 | 7.76 | 2.10% |
| 16% | 201 | 126 | 1.67 | 62.7% | 210.2 | 9.20 | 4.57% | 300 | 183 | 1.68 | 61.0% | 308.0 | 8.02 | 2.67% |
| 17% | 138 | 83 | 1.67 | 60.1% | 138.8 | 0.80 | 0.58% | 237 | 149 | 1.70 | 62.9% | 253.7 | 16.75 | 7.07% |
| 18% | 113 | 70 | 1.68 | 61.9% | 117.7 | 4.72 | 4.18% | 180 | 108 | 1.69 | 60.0% | 182.1 | 2.12 | 1.18% |
| 19% | 87 | 53 | 1.66 | 60.9% | 87.9 | 0.86 | 0.99% | 136 | 85 | 1.70 | 62.5% | 144.5 | 8.48 | 6.24% |
| 20% | 62 | 40 | 1.74 | 64.5% | 69.7 | 7.65 | 12.34% | 108 | 69 | 1.72 | 63.9% | 118.8 | 10.81 | 10.01% |
| 21% | 44 | 31 | 1.74 | 70.5% | 54.0 | 10.01 | 22.75% | 78 | 51 | 1.73 | 65.4% | 88.3 | 10.34 | 13.26% |
| 22% | 30 | 20 | 1.87 | 66.7% | 37.3 | 7.31 | 24.37% | 60 | 40 | 1.77 | 66.7% | 70.8 | 10.81 | 18.02% |
| 23% | 24 | 17 | 1.85 | 70.8% | 31.4 | 7.37 | 30.71% | 41 | 26 | 1.95 | 63.4% | 50.7 | 9.70 | 23.66% |
| 24% | 16 | 11 | 1.84 | 68.8% | 20.2 | 4.20 | 26.25% | 33 | 21 | 2.05 | 63.6% | 43.0 | 9.98 | 30.24% |
| 25% | 11 | 8 | 1.81 | 72.7% | 14.5 | 3.45 | 31.36% | 27 | 19 | 2.04 | 70.4% | 38.7 | 11.73 | 43.44% |

**Table 13.** The payoff discrepancies *θ* that maximise **ROI** per season (in yellow shading), based on **AH bets** and for both average (left) and maximum (right) market odds. The optimal *θ* discrepancy is chosen over all *θ* that generate at least 30 bets in a single season.

| | Average market odds | | | | | | | | Maximum market odds | | | | | | | |
|---|---|---|---|---|---|---|---|---|---|---|---|---|---|---|---|---|
| **Season** | **θ** | **Bets** | **Bets won** | **Odds** | **Win Rate** | **Returns** | **Profit** | **ROI** | **θ** | **Bets** | **Bets won** | **Odds** | **Win Rate** | **Returns** | **Profit** | **ROI** |
| 2006/07 | 7% | 66 | 46 | 1.6 | 69.7% | 74.9 | 8.945 | 13.55% | 8% | 80 | 57 | 1.6 | 71.3% | 91.4 | 11.385 | 14.23% |
| 2007/08 | 8% | 71 | 47 | 1.5 | 66.2% | 72.2 | 1.165 | 1.64% | 10% | 70 | 47 | 1.5 | 67.1% | 72.7 | 2.745 | 3.92% |
| 2008/09 | 12% | 30 | 22 | 1.8 | 73.3% | 38.8 | 8.78 | 29.27% | 14% | 31 | 21 | 1.9 | 67.7% | 40.9 | 9.88 | 31.87% |
| 2009/10 | 11% | 52 | 33 | 1.5 | 63.5% | 49.6 | -2.39 | -4.60% | 16% | 31 | 22 | 1.4 | 71.0% | 31.6 | 0.61 | 1.97% |
| 2010/11 | 11% | 43 | 28 | 1.8 | 65.1% | 51.8 | 8.75 | 20.35% | 14% | 33 | 22 | 1.8 | 66.7% | 39.0 | 6.01 | 18.21% |
| 2011/12 | 10% | 51 | 29 | 1.8 | 56.9% | 51.7 | 0.74 | 1.45% | 14% | 34 | 20 | 1.8 | 58.8% | 36.0 | 1.98 | 5.82% |
| 2012/13 | 9% | 55 | 40 | 1.6 | 72.7% | 65.6 | 10.55 | 19.18% | 13% | 30 | 23 | 1.5 | 76.7% | 35.4 | 5.415 | 18.05% |
| 2013/14 | 10% | 58 | 38 | 1.7 | 65.5% | 64.0 | 5.96 | 10.28% | 14% | 33 | 22 | 1.7 | 66.7% | 37.9 | 4.92 | 14.91% |
| 2014/15 | 8% | 64 | 39 | 1.8 | 60.9% | 68.7 | 4.68 | 7.31% | 9% | 72 | 46 | 1.8 | 63.9% | 83.4 | 11.41 | 15.85% |
| 2015/16 | 8% | 109 | 68 | 1.6 | 62.4% | 107.2 | -1.76 | -1.61% | 11% | 87 | 55 | 1.7 | 63.2% | 91.7 | 4.745 | 5.45% |
| 2016/17 | 15% | 33 | 21 | 1.6 | 63.6% | 33.4 | 0.405 | 1.23% | 17% | 33 | 21 | 1.6 | 63.6% | 34.2 | 1.18 | 3.58% |
| 2017/18 | 14% | 41 | 25 | 1.7 | 61.0% | 43.3 | 2.29 | 5.59% | 17% | 31 | 20 | 1.8 | 64.5% | 36.3 | 5.305 | 17.11% |
| 2018/19 | 2% | 267 | 163 | 1.7 | 61.0% | 272.5 | 5.485 | 2.05% | 4% | 256 | 156 | 1.7 | 60.9% | 266.6 | 10.59 | 4.14% |
| Overall | 7.45% | 940 | 599 | 1.66 | 63.7% | 993.6 | 53.6 | 5.70% | 10.04% | 821 | 532 | 1.68 | 64.8% | 897.2 | 76.175 | 9.28% |

**Table 14.** The payoff discrepancies *θ* that maximise **profit** per season (in yellow shading), based on **AH bets** and for both average (left) and maximum (right) market odds. The optimal *θ* discrepancy is chosen over all *θ* that generate at least 30 bets in a single season.

| | Average market odds | | | | | | | | Maximum market odds | | | | | | | |
|---|---|---|---|---|---|---|---|---|---|---|---|---|---|---|---|---|
| **Season** | **θ** | **Bets** | **Bets won** | **Odds** | **Win Rate** | **Returns** | **Profit** | **ROI** | **θ** | **Bets** | **Bets won** | **Odds** | **Win Rate** | **Returns** | **Profit** | **ROI** |
| 2006/07 | 1% | 227 | 151 | 1.6 | 66.5% | 244.2 | 17.22 | 7.59% | 1% | 308 | 204 | 1.7 | 66.2% | 339.5 | 31.46 | 10.21% |
| 2007/08 | 8% | 71 | 47 | 1.5 | 66.2% | 72.2 | 1.165 | 1.64% | 8% | 103 | 68 | 1.6 | 66.0% | 106.6 | 3.64 | 3.53% |
| 2008/09 | 9% | 67 | 46 | 1.8 | 68.7% | 80.9 | 13.89 | 20.73% | 6% | 168 | 102 | 1.8 | 60.7% | 188.5 | 20.47 | 12.18% |
| 2009/10 | 12% | 39 | 25 | 1.5 | 64.1% | 37.1 | -1.89 | -4.85% | 16% | 31 | 22 | 1.4 | 71.0% | 31.6 | 0.61 | 1.97% |
| 2010/11 | 11% | 43 | 28 | 1.8 | 65.1% | 51.8 | 8.75 | 20.35% | 0% | 356 | 220 | 1.7 | 61.8% | 376.7 | 20.68 | 5.81% |
| 2011/12 | 10% | 51 | 29 | 1.8 | 56.9% | 51.7 | 0.74 | 1.45% | 14% | 34 | 20 | 1.8 | 58.8% | 36.0 | 1.98 | 5.82% |
| 2012/13 | 9% | 55 | 40 | 1.6 | 72.7% | 65.6 | 10.55 | 19.18% | 6% | 154 | 104 | 1.6 | 67.5% | 163.0 | 9.02 | 5.86% |
| 2013/14 | 4% | 172 | 111 | 1.7 | 64.5% | 188.9 | 16.9 | 9.83% | 6% | 161 | 104 | 1.8 | 64.6% | 183.0 | 22.035 | 13.69% |
| 2014/15 | 6% | 95 | 58 | 1.8 | 61.1% | 101.6 | 6.64 | 6.99% | 9% | 72 | 46 | 1.8 | 63.9% | 83.4 | 11.41 | 15.85% |
| 2015/16 | 8% | 109 | 68 | 1.6 | 62.4% | 107.2 | -1.76 | -1.61% | 11% | 87 | 55 | 1.7 | 63.2% | 91.7 | 4.745 | 5.45% |
| 2016/17 | 15% | 33 | 21 | 1.6 | 63.6% | 33.4 | 0.405 | 1.23% | 17% | 33 | 21 | 1.6 | 63.6% | 34.2 | 1.18 | 3.58% |
| 2017/18 | 14% | 41 | 25 | 1.7 | 61.0% | 43.3 | 2.29 | 5.59% | 0% | 369 | 240 | 1.6 | 65.0% | 382.6 | 13.565 | 3.68% |
| 2018/19 | 2% | 267 | 163 | 1.7 | 61.0% | 272.5 | 5.485 | 2.05% | 2% | 318 | 197 | 1.7 | 61.9% | 330.5 | 12.515 | 3.94% |
| Overall | 5.57% | 1270 | 812 | 1.66 | 63.9% | 1350.4 | 80.385 | 6.33% | 5.55% | 2194 | 1403 | 1.69 | 63.9% | 2347.3 | 153.31 | 6.99% |

### *5.3.1 Odds of bets simulated*

When it comes to the bets simulated, the 1X2 bets tend to average odds greater than 3 which suggests that the model tends to recommend bets on outsiders; a behaviour that is consistent with previous studies including the original pi-rating (Constantinou & Fenton, 2013; Constantinou, 2018). Conversely, the AH bets tend to be simulated on favourite outcomes with average season odds typically ranging between 1.6 and 1.8. However, it is important to note that an issue with the AH odds retrieved from www.football-data.co.uk is that they do not always represent the odds associated with the handicap that maximises the uniformity of the AH distribution, as discussed in Section 2. For example, the AH odds for seasons 2009/10 and 2010/11 appear to be predominantly based on ±0 AH; i.e., no handicap, with the outcome of draw eliminated. Examples of this issue can also be viewed in Table A2; e.g., refer to the imbalanced AH odds for dates 14/08, 11/09 and 27/11.

According to Table 15, at least part of the AH odds of the first five seasons do not reflect the standard AH outcome, whereas the eight most recent seasons appear to be correctly based on the standard AH outcome that aims to make the competition equal. Results from predictive accuracy and profitability suggest that there is no meaningful difference between the first five and the last eight seasons. Therefore, we have no reason to assume that this might have influenced the overall conclusions. Finally, the preference of the model to bet on favourite AH outcomes remains consistent across all 13 seasons. This outcome is also discussed in Section 6, point *ii*.

**Table 15.** The mean average and mean maximum AH odds for each of the 13 seasons.

| | Average odds | | Maximum odds | |
|---|---|---|---|---|
| **Season** | **HT** | **AT** | **HT** | **AT** |
| 2006/07 | 1.89 | 1.97 | 1.95 | 2.05 |
| 2007/08 | 1.92 | 2.01 | 2.00 | 2.09 |
| 2008/09 | 1.85 | 2.30 | 1.94 | 2.50 |
| 2009/10 | 2.08 | 3.01 | 2.24 | 3.38 |
| 2010/11 | 1.87 | 2.24 | 1.94 | 2.38 |
| 2011/12 | 1.93 | 1.94 | 1.99 | 2.01 |
| 2012/13 | 1.93 | 1.95 | 1.99 | 2.01 |
| 2013/14 | 1.93 | 1.94 | 2.00 | 2.01 |
| 2014/15 | 1.92 | 1.95 | 1.98 | 2.01 |
| 2015/16 | 1.94 | 1.93 | 1.99 | 1.99 |
| 2016/17 | 1.95 | 1.93 | 2.01 | 1.99 |
| 2017/18 | 1.95 | 1.93 | 2.00 | 1.99 |
| 2018/19 | 1.96 | 1.94 | 2.03 | 2.00 |

A possible limitation here is that, while the AH market offers multiple handicaps for each match, this study has only considered one handicap per match. However, it is reasonable to assume that the results presented in this paper approximate the overall AH market. This is because when the model suggests a bet on team $X$ for a given handicap, then we should expect the model to suggest a bet on team $X$ regardless the handicap, since any handicap must remain faithful to the expected goal difference of the match, which determines $\theta$.

### *5.3.2 Betting stake adjustments*

Fig 5 presents the cumulative profit generated over eight different betting scenarios that represent the combinations of the following betting options: a) optimising for maximum ROI or profit, b) optimising $\theta$ per season or across all seasons, and c) simulating 1X2 or AH bets. The results illustrate how the difference in profit and ROI evolves across the 13 seasons between 1X2 and AH bets. While AH bets generate considerably lower profit and ROI, the profitability is much less volatile than 1X2 bets and hence, it is subject to a lower risk of loss which can often be detrimental. For example, note the significant losses for the two best performing scenarios during matches 1900 to 2100, which are both based on 1X2 bets. However, the lower risk of loss also limits profits.

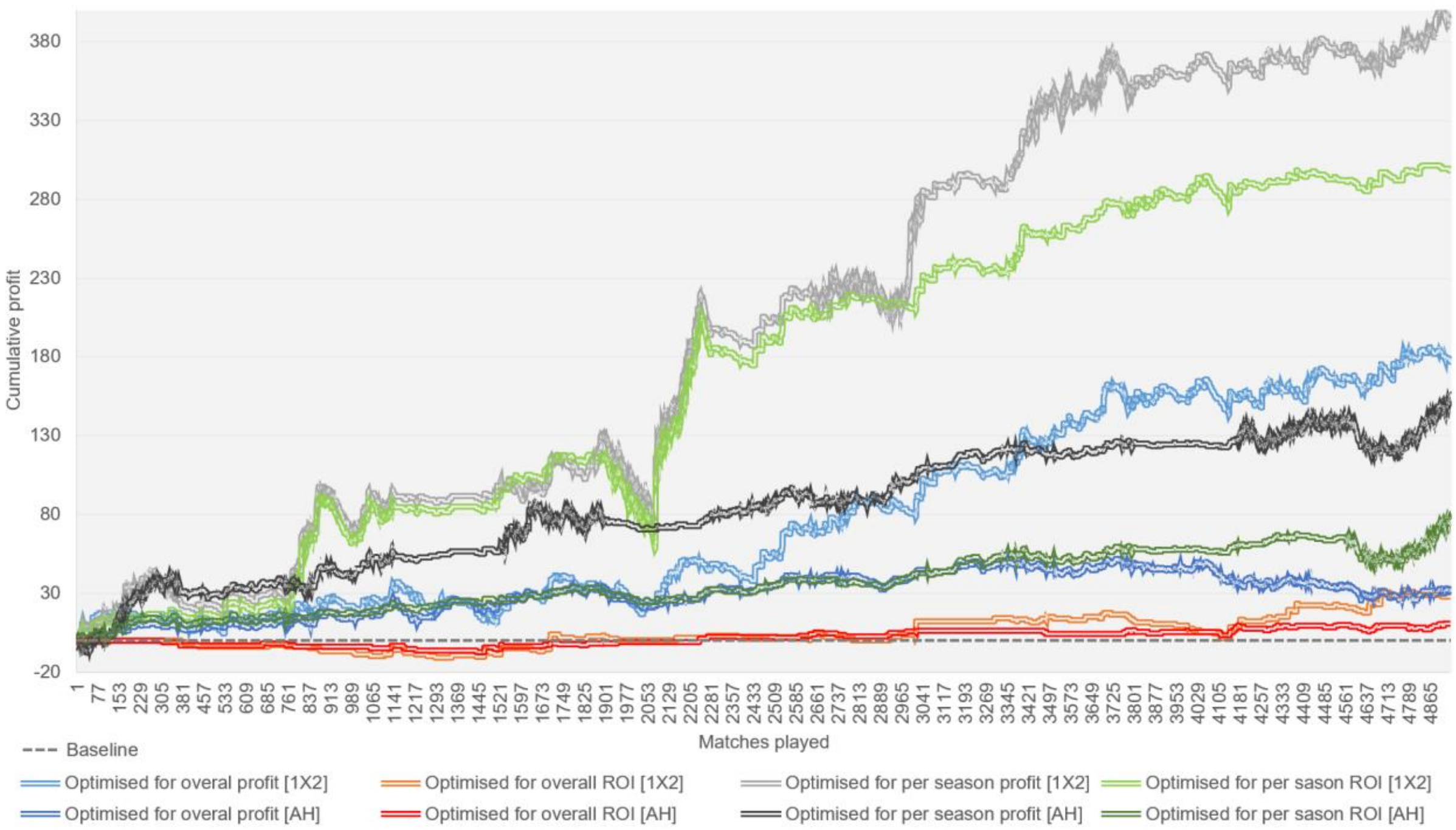

**Figure 5.** Cumulative profit when the betting procedure is optimised for either profit or ROI, overall or per season, and based on either 1X2 or AH maximum market odds. The results are based on 13 EPL seasons; from 2006/09 to 2018/19. Optimisations for overall profit and ROI, across all 13 seasons, are restricted to $\theta$ discrepancies that generate at least 100 bets over those 13 seasons.

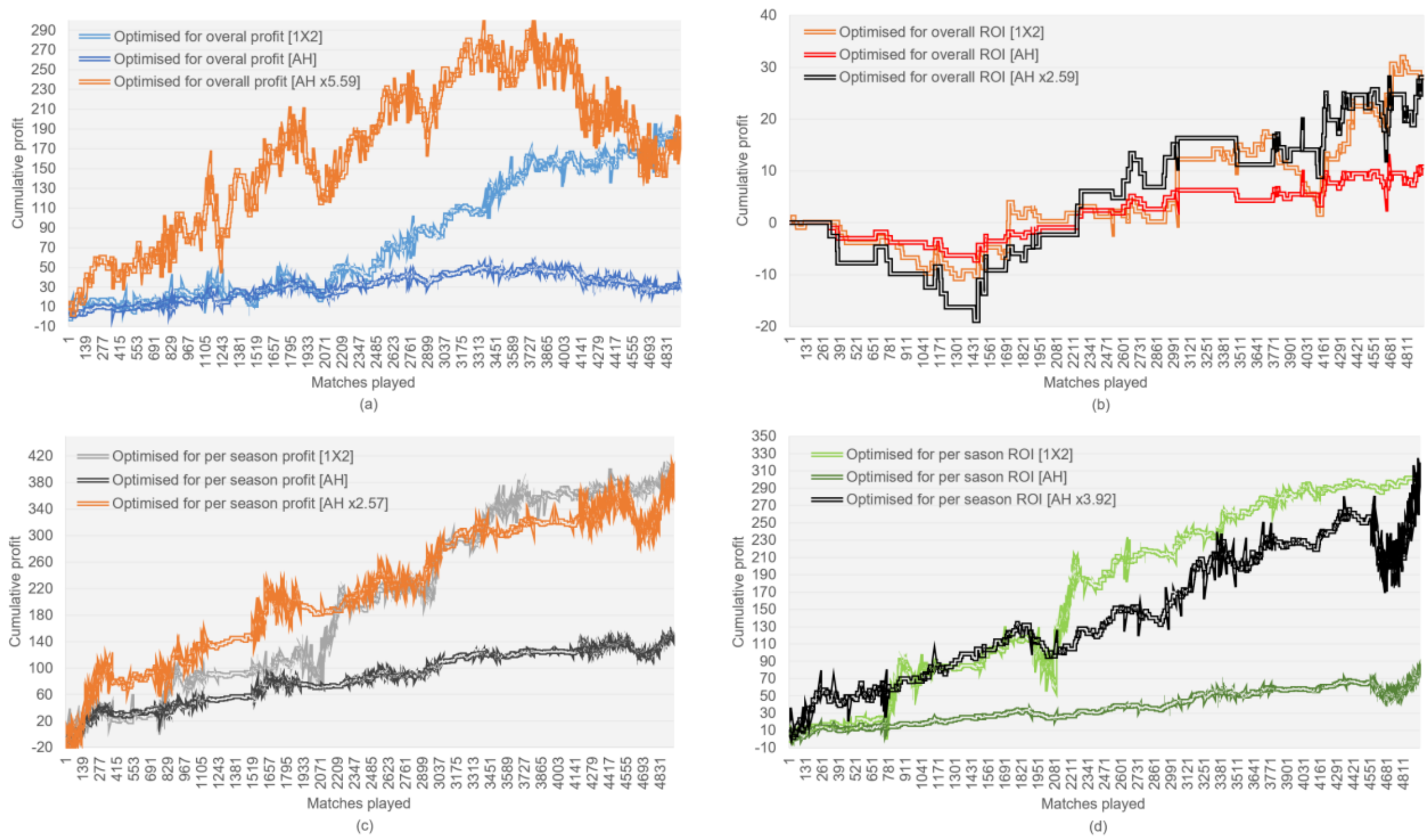

**Figure 6.** Comparing the volatility of profits when the stakes of AH bets is increased by as much required for the cumulative profit to match that of 1X2 bets.

A fairer assessment of risk between 1X2 and AH profits would be to simply optimise stakes such that, at the end of the betting period, they both produce the same profit. Fig 6 provides these results by extending the scenarios of Fig 5 to include an additional betting scenario in which AH stakes are increased proportional to the difference in cumulative profit between 1X2 and AH bets. For example, in Fig 6a the new AH betting scenario assumes an increase of 5.59 times the stakes of AH bets, in order for the cumulative profit to become equal to that generated by 1X2 bets.

Overall, the graphs suggest that if we want AH bets to generate as much profit as 1X2 bets do, then profit from AH bets will likely be subject to a similar risk of loss as with 1X2 bets. Therefore, while AH is often preferred due to the lower variance of returns, this advantage is rather eliminated when we need to bet proportionally larger to match the expected profit produced by the corresponding 1X2 bets. This outcome is also discussed in Section 6, point $iii$.

## 6. Discussion and Concluding Remarks

This paper presented a model specifically developed for the prediction and assessment of the AH football betting market. The model is based on a modified version of the pi-ratings system which measures the relative scoring ability between teams. The modified pi-ratings are used as an input into a novel BN model that had its graphical structure determined by the temporal assumption $Possession \rightarrow Shots \rightarrow Shots\ on\ Target \rightarrow Goals\ scored$, which captures the natural causal chain of these events via a Beta-Binomial Hybrid BN modelling process. One example of this assumption is that possession occurs before shots (or shots on target) and hence, shots are assumed to be more impactful than possession in terms of determining goals scored.

Using goal scoring data over the last 27 EPL seasons, the modified pi-ratings discovered a strong linear relationship between team rating difference and expected goal difference. However, the linear relationship is oscillatory (refer to Fig 3) and this suggests that goal data alone may be insufficient in completely explaining team ability. Future work will investigate whether factors beyond goals scored could better explain this relationship. For example, in (Constantinou & Fenton, 2017) it was shown that the three teams who were promoted to the EPL, from the English Championship, tend to perform significantly better than the teams they replace. This is an important factor not taken into consideration by the pi-ratings; i.e., the teams are promoted with either an ignorant rating (if it is their first time in the EPL) or with the rating they had when they were last relegated, which clearly underestimates their performance once they return to the EPL.

AH betting is assessed with reference to the traditional 1X2 betting. The assessment is based on both average and maximum market odds and over all possible betting decision thresholds in terms of discrepancy between predicted and offered market odds. Furthermore, the assessment differentiates between betting strategies that are optimised for ROI and betting strategies that are optimised for profit. Key observations include:

i. The previous literature has generally focused on maximum market odds, and this is understandable since professional gamblers aim to maximise payoff. Still, average odds are important because they reveal the expected returns for the average gambler. Moreover, maximum odds are not attainable by everyone since many countries do not allow access to many of the online bookmakers, including exchange-based websites which often offer the best odds (excluding commission). This study shows that the maximum available market odds increase profits by up to four times relative to average odds. Specifically, taking advantage of the maximum market odds can lead to increased profits that range anywhere between 42% (refer to overall profits in Table 13) and 296% (refer to maximised profits in Table 9).

ii. The recommended AH bets tend to be on favourite outcomes with odds that typically average between 1.6 and 1.8 per season. Conversely, the recommended 1X2 bets tend to be on outsider outcomes with odds averaging above 3. The reduction of the problem from a three-state multinomial to a binary distribution (i.e., from 1X2 to AH) explains why the odds move from 1-in-3 to 1-in-2, but not why the recommended bets switch from outsiders to favourites.

iii. AH bets generate lower profit as well as ROI compared to 1X2 bets. Specifically, 1X2 bets are found to generate ~2.5 to ~5.5 times higher profit and ~2.5 to ~4 times higher ROI compared to AH bets (refer to Fig 6). For this reason, returns from AH bets tend to be considerably less volatile and subject to a lower risk of loss. While this outcome is in agreement with (Hassanniakalager & Newall, n.d.), this presumed advantage of AH betting is flawed. This is because, when the betting stakes of AH bets are increased proportional to the difference in cumulative profit between 1X2 and AH bets, the variance of profit from AH bets increases towards the variance of profit from 1X2 bets. This implies that, when aiming for the same profit at the end of the same period of time, AH bets are not necessarily less risky than 1X2 bets.

iv. Past studies often focus on a single football season, and profitability tends to be reported based on the betting decision threshold that maximises ROI under the assumption that the optimal betting decision threshold remains invariant between seasons. However, the results in this paper show that the optimal betting decision threshold varies dramatically between seasons, despite predictive accuracy being consistent across the 13 seasons, and this applies to both 1X2 and AH bets; albeit to a lower degree for AH bets.

    This implies that the profitability presented in Tables 10, 11, 13, and 14 is *not* a realistic expectation of real-world performance. This is because the optimal betting decision threshold is not consistent between seasons, and the high variance suggest that it is unreasonable to assume we will be able to predict the decision threshold that maximises profit or ROI before a season starts. Therefore, the choice of evaluating football models based on the threshold that maximises profitability in a single football season, which is often the case in the literature, should be discouraged. Moreover, the optimal betting decision threshold is also dependent on whether we would like to maximise ROI or profit. On the other hand, Tables 9 and 12 represent a more realistic expectation of real-world performance, even though it is unlikely that we will follow a static betting decision threshold across these many seasons.

v. Neither profit nor ROI are consistent between seasons, and this applies to both 1X2 and AH bets. While the overall performance of the model is good enough to beat the market, it is still possible for the best possible betting decision threshold to be lossmaking for a whole season (see Tables 10, 11, 13, 14). While this is true for average market odds, the risk is eliminated when we consider maximum odds; though some seasons were barely profitable.

vi. Finally, the results show that choosing to optimise for maximum ROI per season will likely produce undesired results in the long term, and this applies to both 1X2 and AH bets. On the other hand, choosing to optimise for maximum profit (rather than ROI) per season, not only guarantees that the profit is maximised across all seasons, but also often generates a higher overall ROI, across all seasons, compared to the overall ROI generated when optimising for maximum ROI for each season independently. This finding is important since most of the previous studies focus on maximising ROI, often for individual seasons.

Lastly, it is important to note that this paper has considered football data up to season 2018/19. The two subsequent seasons have been partly affected by the COVID-19 pandemic, where many matches were played with fewer or without fans. Relevant studies have shown that this event had an insignificant or a significant negative effect on home advantage (Wunderlich et al., 2021; McCarrick et al., 2021). The model described in this paper does not to consider this event. However, because the model relies on pi-ratings which involve a home and an away rating for each team, it can be easily adjusted to consider such previously unseen events. For example, if we assume that home advantage is not relevant for a particular match, we could consider assigning the 'away' ratings to both teams for that match. Future research works could investigate whether such modelling modifications, that take into consideration the effect of playing in empty stadiums, improve predictive accuracy.

## Acknowledgements

This research was supported by the ERSRC Fellowship project EP/S001646/1 on Bayesian Artificial Intelligence for Decision Making under Uncertainty, by The Alan Turing Institute in the UK, and by Agena Ltd.

# Appendix A: Sample results from betting simulations

**Table A1**. Details of profitability for case in Table 10: Season=2010/11, Odds=Average, Bets=1X2, Optimisation=ROI, and $\theta$=10%.

| | | | | Model predictions | | | Average bookmakers' odds | | | Bookmakers' unnormalised prediction | | | Payoff discrepancy | | | Bets simulated | | | Returns from bets | | | |
|---|---|---|---|---|---|---|---|---|---|---|---|---|---|---|---|---|---|---|---|---|---|---|
| **Date** | **HT** | **AT** | **1X2** | **p(1)** | **p(X)** | **p(2)** | **Odds(1)** | **Odds(X)** | **Odds(2)** | **p(1)** | **p(X)** | **p(2)** | **θ(1)** | **θ(X)** | **θ(2)** | **Bet(1)** | **Bet(X)** | **Bet(2)** | **Return(1)** | **Return(X)** | **Return(2)** | **Profit** |
| 14/08/2010 | Aston Villa | West Ham | 1 | 0.62 | 0.22 | 0.16 | 1.96 | 3.30 | 4.03 | 0.51 | 0.30 | 0.25 | 0.11 | -0.08 | -0.09 | 1 | 0 | 0 | 1.96 | 0 | 0 | 0.96 |
| 14/08/2010 | Wigan | Blackpool | 2 | 0.33 | 0.26 | 0.41 | 1.82 | 3.45 | 4.50 | 0.55 | 0.29 | 0.22 | -0.22 | -0.03 | 0.19 | 0 | 0 | 1 | 0 | 0 | 4.5 | 3.5 |
| 26/09/2010 | Wolves | Aston Villa | 2 | 0.25 | 0.25 | 0.50 | 2.83 | 3.25 | 2.50 | 0.35 | 0.31 | 0.40 | -0.10 | -0.06 | 0.10 | 0 | 0 | 1 | 0 | 0 | 2.5 | 1.5 |
| 02/10/2010 | Sunderland | Man United | X | 0.13 | 0.20 | 0.67 | 4.93 | 3.45 | 1.75 | 0.20 | 0.29 | 0.57 | -0.08 | -0.09 | 0.10 | 0 | 0 | 1 | 0 | 0 | 0 | -1 |
| 23/10/2010 | Birmingham | Blackpool | 1 | 0.39 | 0.27 | 0.34 | 1.85 | 3.48 | 4.31 | 0.54 | 0.29 | 0.23 | -0.15 | -0.02 | 0.11 | 0 | 0 | 1 | 0 | 0 | 0 | -1 |
| 24/10/2010 | Liverpool | Blackburn | 1 | 0.73 | 0.17 | 0.10 | 1.66 | 3.64 | 5.43 | 0.60 | 0.27 | 0.18 | 0.12 | -0.11 | -0.08 | 1 | 0 | 0 | 1.66 | 0 | 0 | 0.66 |
| 01/11/2010 | Blackpool | West Brom | 1 | 0.47 | 0.26 | 0.26 | 2.79 | 3.25 | 2.54 | 0.36 | 0.31 | 0.39 | 0.11 | -0.04 | -0.13 | 1 | 0 | 0 | 2.79 | 0 | 0 | 1.79 |
| 10/11/2010 | Man City | Man United | X | 0.25 | 0.25 | 0.50 | 2.57 | 3.22 | 2.75 | 0.39 | 0.31 | 0.36 | -0.14 | -0.06 | 0.14 | 0 | 0 | 1 | 0 | 0 | 0 | -1 |
| 27/11/2010 | Bolton | Blackpool | X | 0.41 | 0.27 | 0.32 | 1.57 | 3.96 | 5.82 | 0.64 | 0.25 | 0.17 | -0.22 | 0.01 | 0.15 | 0 | 0 | 1 | 0 | 0 | 0 | -1 |
| 11/12/2010 | Aston Villa | West Brom | 1 | 0.62 | 0.22 | 0.16 | 2.10 | 3.30 | 3.53 | 0.48 | 0.30 | 0.28 | 0.15 | -0.08 | -0.13 | 1 | 0 | 0 | 2.1 | 0 | 0 | 1.1 |
| 11/12/2010 | Stoke | Blackpool | 2 | 0.41 | 0.27 | 0.32 | 1.62 | 3.86 | 5.41 | 0.62 | 0.26 | 0.18 | -0.21 | 0.01 | 0.14 | 0 | 0 | 1 | 0 | 0 | 5.41 | 4.41 |
| 12/12/2010 | Tottenham | Chelsea | X | 0.25 | 0.25 | 0.51 | 2.84 | 3.28 | 2.49 | 0.35 | 0.30 | 0.40 | -0.10 | -0.06 | 0.10 | 0 | 0 | 1 | 0 | 0 | 0 | -1 |
| 13/12/2010 | Man United | Arsenal | 1 | 0.62 | 0.22 | 0.16 | 1.95 | 3.40 | 3.92 | 0.51 | 0.29 | 0.26 | 0.11 | -0.07 | -0.09 | 1 | 0 | 0 | 1.95 | 0 | 0 | 0.95 |
| 28/12/2010 | Sunderland | Blackpool | 2 | 0.41 | 0.27 | 0.32 | 1.60 | 3.81 | 5.82 | 0.63 | 0.26 | 0.17 | -0.21 | 0.00 | 0.15 | 0 | 0 | 1 | 0 | 0 | 5.82 | 4.82 |
| 28/12/2010 | West Brom | Blackburn | 2 | 0.36 | 0.27 | 0.37 | 1.82 | 3.48 | 4.49 | 0.55 | 0.29 | 0.22 | -0.19 | -0.02 | 0.15 | 0 | 0 | 1 | 0 | 0 | 4.49 | 3.49 |
| 05/01/2011 | Arsenal | Man City | X | 0.63 | 0.21 | 0.15 | 1.94 | 3.49 | 3.86 | 0.52 | 0.29 | 0.26 | 0.12 | -0.07 | -0.11 | 1 | 0 | 0 | 0 | 0 | 0 | -1 |
| 05/01/2011 | Everton | Tottenham | 1 | 0.48 | 0.26 | 0.26 | 2.66 | 3.24 | 2.64 | 0.38 | 0.31 | 0.38 | 0.10 | -0.05 | -0.12 | 1 | 0 | 0 | 2.66 | 0 | 0 | 1.66 |
| 15/01/2011 | West Brom | Blackpool | 1 | 0.33 | 0.26 | 0.41 | 1.78 | 3.68 | 4.48 | 0.56 | 0.27 | 0.22 | -0.23 | -0.01 | 0.19 | 0 | 0 | 1 | 0 | 0 | 0 | -1 |
| 16/01/2011 | Liverpool | Everton | X | 0.63 | 0.22 | 0.16 | 2.20 | 3.22 | 3.43 | 0.45 | 0.31 | 0.29 | 0.17 | -0.09 | -0.14 | 1 | 0 | 0 | 0 | 0 | 0 | -1 |
| 23/01/2011 | Blackburn | West Brom | 1 | 0.56 | 0.24 | 0.21 | 2.22 | 3.27 | 3.24 | 0.45 | 0.31 | 0.31 | 0.11 | -0.07 | -0.10 | 1 | 0 | 0 | 2.22 | 0 | 0 | 1.22 |
| 01/02/2011 | West Brom | Wigan | X | 0.41 | 0.27 | 0.32 | 1.71 | 3.58 | 5.13 | 0.58 | 0.28 | 0.19 | -0.17 | -0.01 | 0.13 | 0 | 0 | 1 | 0 | 0 | 0 | -1 |
| 12/02/2011 | Man United | Man City | 1 | 0.67 | 0.20 | 0.13 | 1.77 | 3.57 | 4.61 | 0.56 | 0.28 | 0.22 | 0.10 | -0.08 | -0.08 | 1 | 0 | 0 | 1.77 | 0 | 0 | 0.77 |
| 12/02/2011 | West Brom | West Ham | X | 0.36 | 0.27 | 0.37 | 1.93 | 3.52 | 3.86 | 0.52 | 0.28 | 0.26 | -0.16 | -0.02 | 0.11 | 0 | 0 | 1 | 0 | 0 | 0 | -1 |
| 20/02/2011 | West Brom | Wolves | X | 0.36 | 0.26 | 0.37 | 1.86 | 3.40 | 4.33 | 0.54 | 0.29 | 0.23 | -0.18 | -0.03 | 0.14 | 0 | 0 | 1 | 0 | 0 | 0 | -1 |
| 26/02/2011 | Wolves | Blackpool | 1 | 0.36 | 0.26 | 0.37 | 1.83 | 3.58 | 4.28 | 0.55 | 0.28 | 0.23 | -0.18 | -0.02 | 0.14 | 0 | 0 | 1 | 0 | 0 | 0 | -1 |
| 19/03/2011 | Man United | Bolton | 1 | 0.83 | 0.12 | 0.06 | 1.40 | 4.41 | 8.54 | 0.71 | 0.23 | 0.12 | 0.11 | -0.11 | -0.06 | 1 | 0 | 0 | 1.4 | 0 | 0 | 0.4 |
| 19/03/2011 | West Brom | Arsenal | X | 0.12 | 0.20 | 0.68 | 4.67 | 3.59 | 1.76 | 0.21 | 0.28 | 0.57 | -0.09 | -0.08 | 0.11 | 0 | 0 | 1 | 0 | 0 | 0 | -1 |
| 09/04/2011 | Wolves | Everton | 2 | 0.25 | 0.25 | 0.50 | 2.55 | 3.25 | 2.80 | 0.39 | 0.31 | 0.36 | -0.14 | -0.06 | 0.14 | 0 | 0 | 1 | 0 | 0 | 2.8 | 1.8 |
| 11/04/2011 | Liverpool | Man City | 1 | 0.55 | 0.24 | 0.21 | 2.59 | 3.21 | 2.76 | 0.39 | 0.31 | 0.36 | 0.16 | -0.07 | -0.15 | 1 | 0 | 0 | 2.59 | 0 | 0 | 1.59 |
| 14/05/2011 | West Brom | Everton | 1 | 0.26 | 0.24 | 0.50 | 2.63 | 3.28 | 2.69 | 0.38 | 0.30 | 0.37 | -0.12 | -0.06 | 0.13 | 0 | 0 | 1 | 0 | 0 | 0 | -1 |
| 22/05/2011 | Bolton | Man City | 2 | 0.33 | 0.26 | 0.41 | 4.92 | 3.72 | 1.69 | 0.20 | 0.27 | 0.59 | 0.13 | -0.01 | -0.19 | 1 | 0 | 0 | 0 | 0 | 0 | -1 |
| 22/05/2011 | Man United | Blackpool | 1 | 0.82 | 0.12 | 0.06 | 1.56 | 4.13 | 5.57 | 0.64 | 0.24 | 0.18 | 0.18 | -0.12 | -0.12 | 1 | 0 | 0 | 1.56 | 0 | 0 | 0.56 |
| 22/05/2011 | Stoke | Wigan | 2 | 0.55 | 0.24 | 0.21 | 2.76 | 3.42 | 2.45 | 0.36 | 0.29 | 0.41 | 0.19 | -0.05 | -0.20 | 1 | 0 | 0 | 0 | 0 | 0 | -1 |
| **TOTAL** | | | | | | | | | | | | | | | | **15** | **0** | **18** | **22.66** | **0** | **25.52** | **15.18** |

**Table A2**. Details of profitability for case in Table 13: Season=2010/11, Odds=Average, Bets=AH, Optimisation=ROI, and $\theta$=11%.

| | | | Goals | | Goal | | Model predictions | | Average bookmakers' | | Bookmakers' unnormalized | | Payoff discrepancy | | Bets simulated | | Returns from bets | | |
|---|---|---|---|---|---|---|---|---|---|---|---|---|---|---|---|---|---|---|---|
| **Date** | **HT** | **AT** | **HT** | **AT** | **difer** | **AH** | **p(1)** | **p(2)** | **Odds(1)** | **Odds(2)** | **p(1)** | **p(2)** | **θ(1)** | **θ(2)** | **Bet(1)** | **Bet(2)** | **Return(1)** | **Return(2)** | **Profit** |
| 14/08/2010 | Wigan | Blackpool | 0 | 4 | -4 | 0 | 0.45 | 0.55 | 1.32 | 3.19 | 0.76 | 0.31 | -0.31 | 0.24 | 0 | 1 | 0 | 3.19 | 2.19 |
| 21/08/2010 | Arsenal | Blackpool | 6 | 0 | 6 | -2 | 0.28 | 0.72 | 1.76 | 2.12 | 0.57 | 0.47 | -0.28 | 0.24 | 0 | 1 | 0 | 0 | -1 |
| 11/09/2010 | Newcastle | Blackpool | 0 | 2 | -2 | 0 | 0.65 | 0.35 | 1.20 | 4.23 | 0.83 | 0.24 | -0.18 | 0.11 | 0 | 1 | 0 | 4.23 | 3.23 |
| 25/09/2010 | West Ham | Tottenham | 1 | 0 | 1 | 0 | 0.51 | 0.49 | 2.54 | 1.47 | 0.39 | 0.68 | 0.11 | -0.19 | 1 | 0 | 2.54 | 0 | 1.54 |
| 23/10/2010 | Birmingham | Blackpool | 2 | 0 | 2 | -0.5 | 0.39 | 0.61 | 1.85 | 2.01 | 0.54 | 0.50 | -0.15 | 0.11 | 0 | 1 | 0 | 0 | -1 |
| 24/10/2010 | Liverpool | Blackburn | 2 | 1 | 1 | -0.75 | 0.69 | 0.31 | 1.85 | 2.02 | 0.54 | 0.50 | 0.15 | -0.19 | 1 | 0 | 1.425 | 0 | 0.425 |
| 30/10/2010 | Man United | Tottenham | 2 | 0 | 2 | -1 | 0.65 | 0.35 | 2.04 | 1.83 | 0.49 | 0.55 | 0.16 | -0.19 | 1 | 0 | 2.04 | 0 | 1.04 |
| 01/11/2010 | Blackpool | West Brom | 2 | 1 | 1 | 0 | 0.64 | 0.36 | 1.99 | 1.81 | 0.50 | 0.55 | 0.14 | -0.20 | 1 | 0 | 1.99 | 0 | 0.99 |
| 10/11/2010 | Man City | Man United | 0 | 0 | 0 | 0 | 0.33 | 0.67 | 1.82 | 1.98 | 0.55 | 0.51 | -0.22 | 0.16 | 0 | 1 | 0 | 1 | 0 |
| 20/11/2010 | Birmingham | Chelsea | 1 | 0 | 1 | 0.75 | 0.36 | 0.64 | 1.88 | 2.00 | 0.53 | 0.50 | -0.17 | 0.14 | 0 | 1 | 0 | 0 | -1 |
| 27/11/2010 | Bolton | Blackpool | 2 | 2 | 0 | 0 | 0.56 | 0.44 | 1.20 | 4.17 | 0.83 | 0.24 | -0.27 | 0.20 | 0 | 1 | 0 | 1 | 0 |
| 11/12/2010 | Aston Villa | West Brom | 2 | 1 | 1 | 0 | 0.80 | 0.20 | 1.50 | 2.49 | 0.67 | 0.40 | 0.13 | -0.20 | 1 | 0 | 1.5 | 0 | 0.5 |
| 11/12/2010 | Stoke | Blackpool | 0 | 1 | -1 | -1 | 0.24 | 0.76 | 2.07 | 1.82 | 0.48 | 0.55 | -0.24 | 0.21 | 0 | 1 | 0 | 1.82 | 0.82 |
| 26/12/2010 | Aston Villa | Tottenham | 1 | 2 | -1 | 0 | 0.57 | 0.43 | 2.22 | 1.64 | 0.45 | 0.61 | 0.12 | -0.18 | 1 | 0 | 0 | 0 | -1 |
| 28/12/2010 | Sunderland | Blackpool | 0 | 2 | -2 | -1 | 0.25 | 0.75 | 2.06 | 1.81 | 0.49 | 0.55 | -0.24 | 0.20 | 0 | 1 | 0 | 1.81 | 0.81 |
| 28/12/2010 | West Brom | Blackburn | 1 | 3 | -2 | -0.5 | 0.36 | 0.64 | 1.82 | 2.05 | 0.55 | 0.49 | -0.19 | 0.15 | 0 | 1 | 0 | 2.05 | 1.05 |
| 29/12/2010 | Chelsea | Bolton | 1 | 0 | 1 | -1.5 | 0.62 | 0.38 | 1.98 | 1.89 | 0.51 | 0.53 | 0.12 | -0.15 | 1 | 0 | 0 | 0 | -1 |
| 01/01/2011 | Man City | Blackpool | 1 | 0 | 1 | -1.5 | 0.37 | 0.63 | 1.83 | 2.03 | 0.55 | 0.49 | -0.17 | 0.13 | 0 | 1 | 0 | 2.03 | 1.03 |
| 05/01/2011 | Arsenal | Man City | 0 | 0 | 0 | -0.5 | 0.63 | 0.37 | 1.94 | 1.93 | 0.52 | 0.52 | 0.12 | -0.15 | 1 | 0 | 0 | 0 | -1 |
| 05/01/2011 | Everton | Tottenham | 2 | 1 | 1 | 0 | 0.65 | 0.35 | 1.90 | 1.92 | 0.53 | 0.52 | 0.12 | -0.17 | 1 | 0 | 1.9 | 0 | 0.9 |
| 15/01/2011 | West Brom | Blackpool | 3 | 2 | 1 | -0.75 | 0.26 | 0.74 | 1.96 | 1.91 | 0.51 | 0.52 | -0.25 | 0.22 | 0 | 1 | 0 | 0.5 | -0.5 |
| 16/01/2011 | Liverpool | Everton | 2 | 2 | 0 | 0 | 0.80 | 0.20 | 1.55 | 2.41 | 0.65 | 0.41 | 0.16 | -0.22 | 1 | 0 | 1 | 0 | 0 |
| 01/02/2011 | West Brom | Wigan | 2 | 2 | 0 | -0.75 | 0.34 | 0.66 | 1.89 | 1.98 | 0.53 | 0.51 | -0.19 | 0.16 | 0 | 1 | 0 | 1.98 | 0.98 |
| 12/02/2011 | Man United | Man City | 2 | 1 | 1 | -0.75 | 0.62 | 0.38 | 1.99 | 1.89 | 0.50 | 0.53 | 0.12 | -0.15 | 1 | 0 | 1.495 | 0 | 0.495 |
| 12/02/2011 | West Brom | West Ham | 3 | 3 | 0 | -0.5 | 0.36 | 0.64 | 1.92 | 1.96 | 0.52 | 0.51 | -0.16 | 0.13 | 0 | 1 | 0 | 1.96 | 0.96 |
| 20/02/2011 | West Brom | Wolves | 1 | 1 | 0 | -0.5 | 0.36 | 0.64 | 1.88 | 2.01 | 0.53 | 0.50 | -0.17 | 0.14 | 0 | 1 | 0 | 2.01 | 1.01 |
| 26/02/2011 | Wolves | Blackpool | 4 | 0 | 4 | -0.5 | 0.36 | 0.64 | 1.83 | 2.05 | 0.55 | 0.49 | -0.18 | 0.15 | 0 | 1 | 0 | 0 | -1 |
| 05/03/2011 | Arsenal | Sunderland | 0 | 0 | 0 | -1 | 0.67 | 0.33 | 1.82 | 2.06 | 0.55 | 0.49 | 0.12 | -0.16 | 1 | 0 | 0 | 0 | -1 |
| 19/03/2011 | Man United | Bolton | 1 | 0 | 1 | -1.25 | 0.70 | 0.30 | 2.01 | 1.87 | 0.50 | 0.53 | 0.20 | -0.23 | 1 | 0 | 0.5 | 0 | -0.5 |
| 19/03/2011 | West Brom | Arsenal | 2 | 2 | 0 | 0.75 | 0.37 | 0.63 | 1.84 | 2.03 | 0.54 | 0.49 | -0.18 | 0.14 | 0 | 1 | 0 | 0 | -1 |
| 03/04/2011 | Fulham | Blackpool | 3 | 0 | 3 | -1 | 0.31 | 0.69 | 1.92 | 1.94 | 0.52 | 0.52 | -0.21 | 0.17 | 0 | 1 | 0 | 0 | -1 |
| 09/04/2011 | Man United | Fulham | 2 | 0 | 2 | -1 | 0.70 | 0.30 | 1.86 | 2.01 | 0.54 | 0.50 | 0.16 | -0.20 | 1 | 0 | 1.86 | 0 | 0.86 |
| 09/04/2011 | Wolves | Everton | 0 | 3 | -3 | 0 | 0.34 | 0.66 | 1.82 | 2.00 | 0.55 | 0.50 | -0.21 | 0.16 | 0 | 1 | 0 | 2 | 1 |
| 10/04/2011 | Blackpool | Arsenal | 1 | 3 | -2 | 1.5 | 0.65 | 0.35 | 1.87 | 2.00 | 0.53 | 0.50 | 0.11 | -0.15 | 1 | 0 | 0 | 0 | -1 |
| 11/04/2011 | Liverpool | Man City | 3 | 0 | 3 | 0 | 0.72 | 0.28 | 1.86 | 1.97 | 0.54 | 0.51 | 0.19 | -0.23 | 1 | 0 | 1.86 | 0 | 0.86 |
| 16/04/2011 | West Brom | Chelsea | 1 | 3 | -2 | 0.75 | 0.36 | 0.64 | 1.95 | 1.94 | 0.51 | 0.52 | -0.16 | 0.13 | 0 | 1 | 0 | 1.94 | 0.94 |
| 07/05/2011 | Tottenham | Blackpool | 1 | 1 | 0 | -1.5 | 0.40 | 0.60 | 1.81 | 2.05 | 0.55 | 0.49 | -0.16 | 0.12 | 0 | 1 | 0 | 2.05 | 1.05 |
| 14/05/2011 | Sunderland | Wolves | 1 | 3 | -2 | 0 | 0.65 | 0.35 | 1.88 | 1.95 | 0.53 | 0.51 | 0.12 | -0.17 | 1 | 0 | 0 | 0 | -1 |
| 14/05/2011 | West Brom | Everton | 1 | 0 | 1 | 0 | 0.34 | 0.66 | 1.89 | 1.94 | 0.53 | 0.52 | -0.19 | 0.14 | 0 | 1 | 0 | 0 | -1 |
| 15/05/2011 | Arsenal | Aston Villa | 1 | 2 | -1 | -1.25 | 0.40 | 0.60 | 1.81 | 2.07 | 0.55 | 0.48 | -0.15 | 0.11 | 0 | 1 | 0 | 2.07 | 1.07 |
| 22/05/2011 | Bolton | Man City | 0 | 2 | -2 | 0.75 | 0.67 | 0.33 | 2.00 | 1.88 | 0.50 | 0.53 | 0.17 | -0.20 | 1 | 0 | 0 | 0 | -1 |
| 22/05/2011 | Man United | Blackpool | 4 | 2 | 2 | -1 | 0.77 | 0.23 | 2.00 | 1.88 | 0.50 | 0.53 | 0.27 | -0.31 | 1 | 0 | 2 | 0 | 1 |
| 22/05/2011 | Stoke | Wigan | 0 | 1 | -1 | 0 | 0.72 | 0.28 | 2.00 | 1.85 | 0.50 | 0.54 | 0.22 | -0.26 | 1 | 0 | 0 | 0 | -1 |
| **TOTAL** | | | | | | | | | | | | | | | **20** | **23** | **20.11** | **31.64** | **8.75** |